\documentclass[conference]{IEEEtran}
\usepackage{textcomp}
\usepackage{stfloats}
\usepackage{url}
\usepackage{verbatim}
\usepackage{amsmath,amssymb,amsfonts}
\usepackage{amsthm}
\usepackage{subfigure}
\usepackage{makecell}
\usepackage{multirow}
\usepackage{enumitem}
\usepackage{subcaption}

\usepackage{algorithmic}

\usepackage{color, colortbl}

\definecolor{Gray}{rgb}{0.9, 0.9, 0.9}

\usepackage[capitalize,noabbrev]{cleveref}

\usepackage{booktabs}
\usepackage{fontawesome}

\usepackage[ruled,linesnumbered]{algorithm2e}

\usepackage[inkscapelatex=false]{svg}
\def\BibTeX{{\rm B\kern-.05em{\sc i\kern-.025em b}\kern-.08em
    T\kern-.1667em\lower.7ex\hbox{E}\kern-.125emX}}
\begin{document}

\title{CAMMSR: Category-Guided Attentive Mixture of Experts for Multimodal Sequential Recommendation
\noindent \thanks{\textbf{*Corresponding Author is Edith C. H. Ngai.}}}

\makeatletter
\newcommand{\linebreakand}{%
  \end{@IEEEauthorhalign}
  \hfill\mbox{}\par
  \mbox{}\hfill\begin{@IEEEauthorhalign}
}
\makeatother

\author{
\IEEEauthorblockN{1\textsuperscript{st} Jinfeng Xu}
\IEEEauthorblockA{
\textit{The University of Hong Kong}\\
Hong Kong SAR, China \\
jinfeng@connect.hku.hk\\
ORCID: 0009-0001-7876-3740}
\and
\IEEEauthorblockN{2\textsuperscript{nd} Zheyu Chen}
\IEEEauthorblockA{
\textit{Beijing Institute of Technology}\\
Beijing, China\\
zheyu.chen@bit.edu.cn\\
ORCID: 0009-0003-5779-3523}
\and
\IEEEauthorblockN{3\textsuperscript{rd} Shuo Yang}
\IEEEauthorblockA{
\textit{The University of Hong Kong}\\
Hong Kong SAR, China \\
shuoyang.ee@gmail.com\\
ORCID: 0000-0003-1638-9623}
\linebreakand
\IEEEauthorblockN{4\textsuperscript{th} Jinze Li}
\IEEEauthorblockA{
\textit{The University of Hong Kong}\\
Hong Kong SAR, China \\
lijinze-hku@connect.hku.hk\\
ORCID: 0009-0002-6749-5442}
\and
\IEEEauthorblockN{5\textsuperscript{th} Hewei Wang}
\IEEEauthorblockA{
\textit{Carnegie Mellon University}\\
Pittsburgh, USA\\
stephenw2026@gmail.com\\
ORCID: 0000-0002-6952-0886}
\and
\IEEEauthorblockN{6\textsuperscript{th} Yijie Li}
\IEEEauthorblockA{
\textit{Carnegie Mellon University}\\
Pittsburgh, USA \\
yijie.li2022@gmail.com\\
ORCID: 0000-0003-2118-2280}
 \linebreakand
\IEEEauthorblockN{7\textsuperscript{th} Jianheng Tang}
\IEEEauthorblockA{
\textit{Peking University}\\
Beijing, China \\
tangentheng@gmail.com\\
ORCID: 0000-0002-4762-5943}
\and
\IEEEauthorblockN{8\textsuperscript{th} Yunhuai Liu}
\IEEEauthorblockA{
\textit{Peking University}\\
Beijing, China \\
yunhuai.liu@pku.edu.cn\\
ORCID: 0000-0002-1180-8078}

\and
\IEEEauthorblockN{9\textsuperscript{th} Edith C. H. Ngai*}
\IEEEauthorblockA{
\textit{The University of Hong Kong}\\
Hong Kong SAR, China \\
chngai@eee.hku.hk\\
ORCID: 0000-0002-3454-8731}
}

\maketitle

\begin{abstract}
The explosion of multimedia data in information-rich environments has intensified the challenges of personalized content discovery, positioning recommendation systems as an essential form of passive data management. Multimodal sequential recommendation, which leverages diverse item information such as text and images, has shown great promise in enriching item representations and deepening the understanding of user interests. However, most existing models rely on heuristic fusion strategies that fail to capture the dynamic and context-sensitive nature of user-modal interactions. In real-world scenarios, user preferences for modalities vary not only across individuals but also within the same user across different items or categories. Moreover, the synergistic effects between modalities—where combined signals trigger user interest in ways isolated modalities cannot—remain largely underexplored.

To this end, we propose CAMMSR, a Category-guided Attentive Mixture of Experts model for Multimodal Sequential Recommendation. Our framework integrates multimodal content from behavioral sequences and timestamps to improve recommendation robustness and contextual alignment. At its core, CAMMSR introduces a category-guided attentive mixture of experts (CAMoE) module, which learns specialized item representations from multiple perspectives and explicitly models inter-modal synergies. This component dynamically allocates modality weights guided by an auxiliary category prediction task, enabling adaptive fusion of multimodal signals. Additionally, we design a modality swap contrastive learning task to enhance cross-modal representation alignment through sequence-level augmentation. Extensive experiments on four public datasets demonstrate that CAMMSR consistently outperforms state-of-the-art baselines, validating its effectiveness in achieving adaptive, synergistic, and user-centric multimodal sequential recommendation.
\end{abstract}


\section{Introduction}
The exponential growth of internet-scale data has made active information seeking and personal data management increasingly challenging for users, who struggle to articulate their long-term preferences and filter heterogeneous, quality-varying multimedia content effectively \cite{xu2025cohesion,he2020lightgcn,xu2025best,guo2024lgmrec,xu2025vi,chen2025hypercomplex}. Within this context, recommendation systems have emerged as a crucial form of passive data management—intelligently filtering irrelevant information and surfacing personalized content by inferring user interests from historical behavioral interactions. Such systems alleviate user burden by transforming raw interaction data into structured, actionable insights, thereby enabling high-quality, automated information management \cite{chen2025squeeze,xu2025nlgcl,xu2026dggvae}. Among these, sequential recommender systems \cite{sun2019bert4rec,kang2018self} stand out by modeling the temporal dynamics of user behavior, thereby providing a scalable mechanism for user-centric data management in dynamic information environments. By capturing evolving preferences from interaction sequences \cite{dang2023ticoserec,du2023frequency}, these systems provide a scalable and sustainable approach to user-centric data management in increasingly information-rich digital environments.

Building upon the foundation of sequential modeling, recent efforts have sought to further enhance the expressiveness and discriminative power of recommender systems by incorporating auxiliary multimodal data signals, such as item descriptions and images, to improve recommendation fidelity and alignment with nuanced user preferences \cite{xu2024mentor, chen2025don}. By leveraging rich multimodal information associated with items—spanning textual metadata, visual content, and categorical attributes—researchers have developed multimodal sequential recommendation models \cite{xu2025enhancing,liu2021noninvasive,wei2023multi,xu2025lobster}. These works aim to construct more fine-grained item representations, enabling a deeper understanding of user intents and interest shifts across sequential interactions \cite{wang2023missrec,bian2023multi,liang2023mmmlp,xu2024sequence}. As a result, multimodal sequential recommender systems have garnered substantial attention from both academic and industrial communities, positioning themselves as a promising pathway toward more intelligent and human-aligned data management.

Despite these advances, existing multimodal sequential models often fall short in dynamically adapting to the context-dependent nature of user-modal interactions. Users' interests in different modalities are inherently dynamic, varying not only across individuals but also within the same user depending on the specific item or situational context \cite{zhang2024m3oe,liang2023mmmlp}. For example, while some users may prioritize visual appearances when browsing fashion items, others—or even the same user in a different session—may rely more heavily on textual descriptions for product specifications or reviews. Such modal preference shifts are often influenced by item category, purchase intent, or other contextual factors \cite{fu2024iisan,dang2023ticoserec}. Compounding this challenge, the synergistic effects between modalities remain largely underexplored in prior work \cite{liang2023mmmlp,xu2024sequence}. A product may fail to attract interest through any single modality alone, yet the combined presence of complementary signals—such as branded text alongside distinctive visual motifs—can trigger significant engagement. Effectively modeling such cross-modal synergies is critical to closing the gap between static multimodal fusion and real-world user decision processes.

To this end, we propose CAMMSR, a Category-guided Attentive Mixture of Experts model for Multimodal Sequential Recommendation. Our framework holistically integrates multimodal content from item sequences and their corresponding timestamps to achieve more robust and context-aware recommendations. At its core, CAMMSR introduces a category-guided attentive mixture of experts (CAMoE) component, which captures specialized item representations and user interests from multi-perspective experts while explicitly modeling inter-modal synergies. This module dynamically allocates modality-specific weights under the guidance of an auxiliary category prediction task, enabling the model to adaptively balance multimodal signals based on item type and user behavior. Furthermore, we design a modality swap contrastive learning objective to enhance cross-modal alignment and exploit latent correlations between modalities through sequence-level augmentation. Extensive experiments on four public real-world datasets validate the superiority of CAMMSR over state-of-the-art baselines, demonstrating its effectiveness in advancing multimodal sequential recommendation toward more adaptive, synergistic, and user-centric data management. In summary, our work makes the following contributions:
\begin{itemize}[leftmargin=*]
    \item We propose a novel CAMMSR model for multimodal sequential recommendation. To the best of our knowledge, we are the first to leverage an auxiliary category prediction task to allocate modality weights to enhance the performance of multimodal sequential recommendation.
    \item We propose a category-guided attentive mixture of experts that efficiently captures informative item representations and user preferences from multiple perspectives while exploring the synergistic interactions between different modalities. It incorporates an auxiliary category prediction task to dynamically and adaptively balance the contribution of information from different modalities.
    \item We further introduce a tailored modality swap contrastive learning task to better exploit correlations between modalities and strengthen the alignment between representations of different modalities.
    \item Comprehensive experiments on four real-world datasets demonstrate the effectiveness of CAMMSR.
\end{itemize}
\section{Related Work}
In this section, we review prior research relevant to our work, focusing on both traditional sequential recommendations and multimodal sequential recommendations.

\subsection{Traditional Sequential Recommendation}
Sequential recommendation addresses the fundamental task of predicting a user's next interaction item by analyzing their historical behavior sequence. Early foundational works primarily concentrated on developing specialized sequential encoders to effectively model temporal dependencies within user interaction data. For instance, GRU4Rec \cite{hidasi2015session} pioneered the use of gated recurrent units (RNNs) to capture short-term sequential patterns, while Caser \cite{tang2018personalized} introduced convolutional neural networks (CNNs) to model both point-level and union-level sequential behaviors through horizontal and vertical filters. SASRec \cite{kang2018self} significantly advanced the field by adopting a unidirectional self-attention mechanism to weight relevant items across the entire sequence, enabling the capture of long-range dependencies. Building on transformer architectures, BERT4Rec \cite{sun2019bert4rec} further enhanced sequential modeling through bidirectional self-attention, allowing the model to leverage contextual information from both past and future interactions within masked sequences.

Recognizing the critical role of temporal dynamics in user interest evolution, subsequent research has systematically integrated time-aware mechanisms to refine sequential recommendation. TiSASRec \cite{li2020time} explicitly incorporates relative time intervals between interactions into the self-attention calculation, enabling the model to adaptively weight items based on both content and temporal proximity. MEANTIME \cite{cho2020meantime} extends this concept by introducing multiple granularities of time embeddings—including absolute, relative, and categorical temporal features—to more comprehensively represent temporal context. More recently, TGCL4SR \cite{zhang2024temporal} employs temporal graph contrastive learning with carefully designed time perturbation augmentations to strengthen the encoding of temporal transition patterns and improve robustness to temporal noise. Parallel developments include FEARec \cite{du2023frequency}, which introduces a novel frequency ramp structure combining time-domain and frequency-domain self-attention modules to simultaneously capture both low-frequency stable preferences and high-frequency dynamic interests. Similarly, TiCoSeRec \cite{dang2023ticoserec} proposes five distinct time-interval-based sequence augmentation strategies to normalize the temporal distribution of user behaviors, coupled with contrastive learning objectives to enhance temporal representation learning and generalization performance.
Despite these advancements, the aforementioned methods focus solely on behavior sequences and neglect the potential of multimodal data. As a result, they fail to fully utilize multimodal features that can provide richer representations of user interests.

\subsection{Multimodal Sequential Recommendation}
Multimodal sequential recommendation has established itself as a promising paradigm that harnesses rich multimodal signals from items—spanning visual, textual features—to model complex user preferences and sequential behaviors, thereby substantially enhancing recommendation quality \cite{yuan2023go,ye2024harnessing,liang2023mmmlp,li2023text,ji2023online}. A central thrust of existing research lies in the integration of these heterogeneous modalities to construct more expressive and context-aware item representations. Early efforts, such as MV-RNN \cite{cui2018mv}, explored straightforward fusion strategies—including element-wise addition, concatenation, and feature reconstruction—to merge multimodal inputs. UniSRec \cite{hou2022towards} advanced this line by adopting a Mixture-of-Experts (MoE) architecture to facilitate semantic transfer from textual representations, which are subsequently fused with ID embeddings via parametric weightings, enabling cross-domain semantic alignment.

Subsequently, researchers have developed more sophisticated and adaptive fusion mechanisms to improve both the flexibility and scalability of multimodal integration. MISSRec \cite{wang2023missrec} introduced a lightweight, dynamic fusion module that adaptively recalibrates user attention over modalities via efficient gating networks, enabling fine-grained modality weighting. MMSR \cite{hu2023adaptive} leveraged heterogeneous graph neural networks to model structural dependencies among modalities, supporting flexible cross-modal interaction through a dedicated message-passing scheme. Further advancing the MoE paradigm, M3SRec \cite{bian2023multi} incorporated both modality-specific and cross-modal MoE modules, allowing the model to capture specialized and shared patterns across modalities for more comprehensive representation learning. In a complementary direction, TedRec \cite{xu2024sequence} applied the Fast Fourier Transform (FFT) to project ID and text embedding sequences into the frequency domain, achieving efficient sequence-level semantic fusion while reducing high-frequency noise. Most recently, FindRec \cite{wang2025findrec} introduced a theoretically grounded approach based on Stein kernels to ensure cross-modal consistency, formalizing multimodal alignment through an Integrated Information Coordination Module that enforces statistical coherence across modality-specific manifolds.

While these methods demonstrate progress, most of them overlook synergistic effects between different modalities. To this end, we propose category-guided attentive MoE with tailored modality swap contrastive to effectively mine modality features and leverage synergistic effects between different modalities.
\section{Preliminary}
Multimodal sequential recommendation seeks to leverage multimodal item data and users' past behaviors to provide personalized suggestions for their next interaction. Let $\mathcal{U}$ denote the user set and $\mathcal{I}$ denote the item set. For the multimodal scenario, each item $i \in \mathcal{I}$ represented as $i = \langle i_{id}, i_{t}, i_{v} \rangle$, where $i_{id}$ denotes ID embedding for item $i$, while $i_{t}$ and $i_{v}$ denote to the associated text and image content of item $i$, respectively. Then, the historical behavior sequence $S$ of user $u \in \mathcal{U}$ represented as $S = \{i_1, i_2, ..., i_{|S|}\}$, where $i_n \in \mathcal{I}$ denotes the
interacted item at the $n$-th time step. The timestamp sequence is denoted as $T = \{t_1, t_2,...,t_{|S|}\}$.

\textbf{Problem Definition: } Given a user $u$ and $u$’s historical behavior sequence $S$ with multimodal item information, the objective of multimodal sequential recommendation is to predict an item $i$ that the user $u$ might
be interested in at the $(|S|$$+$$1)$-th time step. This can be formulated
as finding the item $i$ that maximizes the conditional probability
given the sequence of interactions:
\begin{equation}
i=\operatorname{argmax}_{i \in \mathcal{I}} P\left(i_{|S|+1}=i \mid S\right).
\end{equation}

The key notations are summarized in Table~\ref{tab:notation}.

\begin{table}[!t]
    \centering
     \caption{Key notations used in this paper.}
    \label{tab:notation}
    % \vskip -0.1in
 \vskip -0.05in
    % \small
\resizebox{\linewidth}{!}{
    \begin{tabular}{c|c}
     \toprule
         Notation& Description\\
         \midrule
         $\mathcal{I}$ & Item set\\
         $\mathcal{U}$ & User set\\
         $S$ & Historical behavior sequence\\         
         $T$ & Timestamp sequence\\
         $\mathbf{E}_m$ & Entire representation for modality $m$\\
         $\bar{S}^m$& Enriched item sequential representations\\
         $\mathbf{Z}^m$& Final item sequential representations\\
          \midrule
         $f_{i_{id}}$& ID representation of item $i$\\
         $f_{i_t}$& Textual representation of item $i$\\
         $f_{i_v}$& Visual representation of item $i$\\
         $\mathbf{e}_{i_{m}}$& Representation of item $i$ for modality $m$\\
         $\hat{\mathbf{e}}_{i_m}$& Revised representation of item $i$ for modality $m$\\
         $\overline{\mathbf{e}}_{i_m}$ & Final representation of item $i$ for modality $m$\\
         $\hat{y}_c^m$& Probability to class $c$\\
         $d_i^m$& Difference between true classification and predicted labels\\
         $w_i^m$& Modality weight for item $i$\\
         $\hat{y}_i$& Final predicted score\\ 
          % \midrule
        %  $\operatorname{Expert}_{*}(\cdot)$& Experts \\
        %  $\operatorname{FFN}(\cdot)$ & Feed-forward networks\\
        %  $\operatorname{Softmax}(\cdot)$& Softmax function\\
        %  $\operatorname{Transformer}(\cdot)$& Transformer model\\
        %  $\operatorname{Dropout}(\cdot)$& Random dropout operation\\
        %  $\operatorname{LayerNorm}(\cdot)$& Layer normalization operation\\
        %  $\operatorname{DyT}(\cdot)$& Dynamic Tanh operation\\
        %  $\operatorname{Swap}(\cdot)$& Random swap operation \\
        %  $\operatorname{Sim}(\cdot)$& Consine similarity\\
        \bottomrule
    \end{tabular}
    }

    \vskip -0.1in
\end{table}

\section{Methodology}
In this section, we introduce the technical details of our
CAMMSR\footnote{Code can be found in the \textbf{Supplemental Material}.}. As illustrated in Figure~\ref{fig:overview}, CAMMSR consists of four main components: Item Representation Initialization, Category-Guided Attentive Mixture of Experts, User Interest Learning, and Modality Swap Contrastive Learning.

\begin{figure*}
    \centering
     % \vskip -0.05in
    \includegraphics[width=1\linewidth]{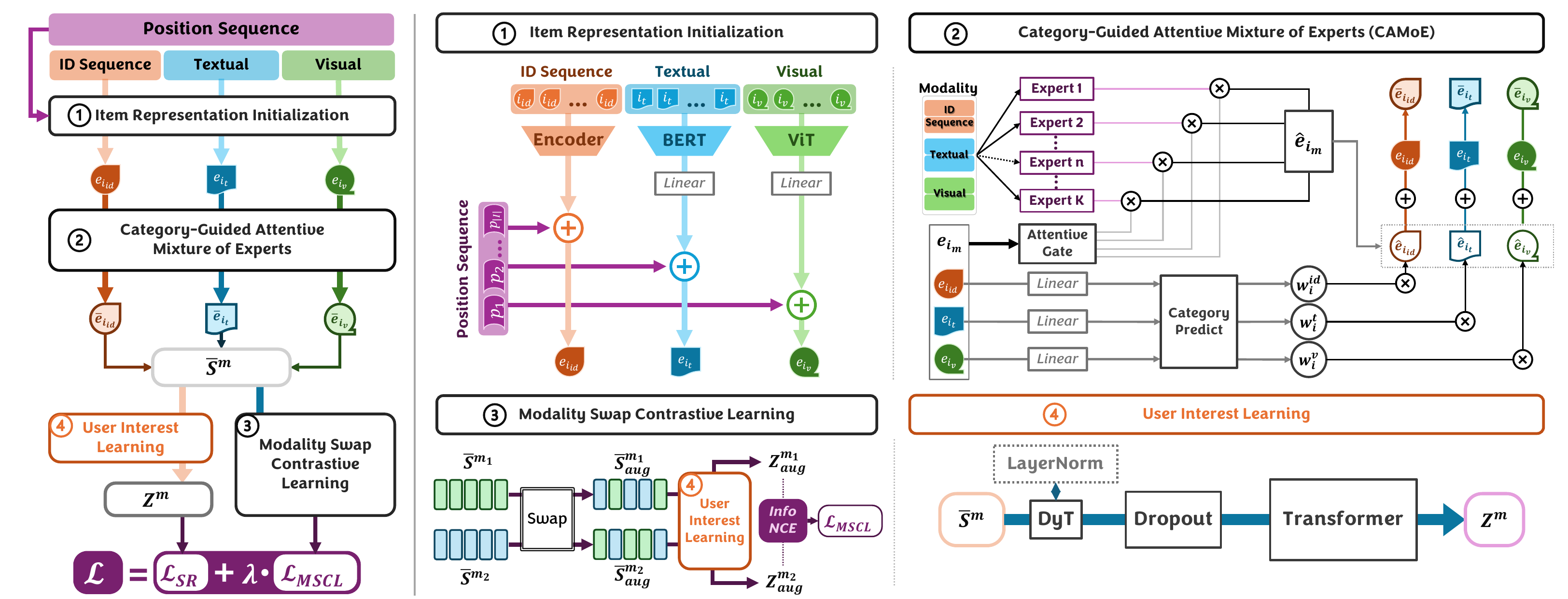}
     % \vskip -0.1in
    \caption{The overall framework of CAMMSR. The left-hand side provides a description of the procedure, while the right-hand side details each component. 1) Item Representation Initialization initializes item multimodal representations using a pre-trained extractor with positional sequences. 2) Category-Guided Attentive Mixture of Experts leverages an additional category prediction task and attention mechanisms to effectively allocate modality weights. 3) Modality Swap Contrastive Learning constructs augmented contrastive views through modality swap operations and applies contrastive learning to enhance the alignment between different modality representations. 4) User Interest Learning models user behavior sequences with a Transformer-based encoder and explores DyT as an alternative to LayerNorm in Transformers.}
     \vskip -0.1in
    \label{fig:overview}
\end{figure*}

\subsection{Item Representation Initialization}
Multimodal information contains rich semantic features \cite{xu2024mentor,zhou2023tale,xu2025survey}, effectively characterizing the items. Therefore, we obtain item representations from three modalities in widely used real-world recommendations: ID, Textual, and Visual. 

\noindent \textbf{ID modality: } We initialize an ID representation $f_{i_{id}} \in \mathbb{R}^{{d}}$ for each item $i$, where $d$ is the dimensionality of the ID modality. 

\noindent \textbf{Textual modality: } Following previous studies \cite{zhang2025hierarchical,bian2023multi}, we apply a widely used pre-trained BERT \cite{devlin2019bert} to extract textual features to capture user preference from textual semantics. We concatenate all textual descriptions $i_{[\text{Words}]}$ with a special symbol $i_{[\text{CLS}_t]}$ for each item $i$. Since $i_{[\text{CLS}_t]}$ can convey the semantics of the whole sentence, we use the embedding of $i_{[\text{CLS}_t]}$ to represent the text features. Here, we input the combined sentence into pre-trained BERT to obtain the textual feature as follows:
\begin{equation}
f_{i_{t}}=\operatorname{BERT}(\operatorname{Con}(i_{[\text{CLS}_t]},i_{[\text{Words}]})),
\end{equation}
where $f_{i_{t}} \in \mathbb{R}^{d_t}$ is the final hidden embedding for $i_{[\text{CLS}_t]}$, and $\operatorname{Con}(\cdot,\cdot)$ denotes the concatenation operation.

\noindent \textbf{Visual modality: } Following previous studies \cite{zhang2025hierarchical,bian2023multi},  we apply a widely used pre-trained ViT \cite{dosovitskiy2020image} to extract visual features to capture user preference from visual semantics. We divide the image for each item $i$ into several patches and then transform these patches into a sequence $i_{[\text{Patches}]}$ with a special symbol $i_{[\text{CLS}_v]}$. Here, we input the combined sentence into pre-trained ViT to obtain the visual feature as follows:
\begin{equation}
f_{i_{v}}=\operatorname{ViT}(\operatorname{Con}(i_{[\text{CLS}_v]},i_{[\text{Patches}]})),
\end{equation}
where $f_{i_{v}} \in \mathbb{R}^{d_v}$ is the final hidden embedding for $i_{[\text{CLS}_v]}$.

To align the dimensionality of textual and visual modalities with ID modality, we adopt a simple linear projection to change their dimensions, formally:
\begin{equation}
x_{i_{t}}=\mathbf{W}_{t} f_{i_{t}} + \mathbf{b}_{t}, \quad x_{i_{v}}=\mathbf{W}_{v} f_{i_{v}} + \mathbf{b}_{v},
\end{equation}
where $\mathbf{W}_{t} \in \mathbb{R}^{d_{t} \times d}$, $\mathbf{W}_{v} \in \mathbb{R}^{d_{v} \times d}$, $\mathbf{b}_t \in \mathbb{R}^{d}$, and $\mathbf{b}_v \in \mathbb{R}^{d}$ are trainable parameters, while $f_{i_{t}}$ and $f_{i_{v}}$ are frozen.

Moreover, position information plays a crucial role in sequence modeling within sequential recommendation. To incorporate position information, we introduce the corresponding positional embedding $p_i \in \mathbb{R}^{d}$ to item representation for all modalities: $\mathbf{e}_{i_{m}} = x_{i_{m}} + p_i$, where $m \in \{id, t, v\}$ denotes the modality. The entire representations for modality $m$ are represented as $\mathbf{E}_m \in \mathbb{R}^{|\mathcal{I}| \times d}$.

\subsection{Category-Guided Attentive Mixture of Experts}
Users' interest in multimodal information is dynamic and not equally weighted. Different users place varying levels of importance on different modalities, and even the same user may prioritize different modalities for different items. For example, the visual modality plays a dominant role when purchasing beauty makeup, whereas the textual modality is more instructive when buying games. To model users' interest in recommendations derived from multimodal information and to allocate appropriate weights to different modalities, we propose the Category-Guided Attentive Mixture of Experts (CAMoE). Specifically, CAMoE further leverages an additional category prediction task to guide the allocation of modality weights in a more effective manner by adopting attention mechanisms to learn the weights of different modalities. Previous studies \cite{bian2023multi,xu2024mome,zhang2024m3oe,ma2018modeling,cai2024survey} have demonstrated the effectiveness of the Mixture of Experts (MoE) framework across various recommendation scenarios. This is attributed to its ability to enhance the learning capacity and flexibility of recommendation models by capturing specialized item representations and user preferences from multiple perspectives. 
% Our proposed CAMoE can be divided into two components: Attentive Mixture of Experts and Category-Guided Fusion. 

Existing studies \cite{bian2023multi,hou2022towards} typically process each modality's feature independently and then apply weighting mechanisms. However, this paradigm often overlooks the synergistic effects between different modalities. For example, marketing information such as discounts or brands described in the textual modality may struggle to directly persuade users to purchase an item, and visual elements like anime patterns alone may also fail to directly trigger purchases. However, discounted items featuring specific anime patterns frequently lead to purchase surges. To this end, we incorporate valuable information from other modalities into the experts corresponding to each modality. Specifically, given the ID, text, and image representations of an item $\mathbf{e}_{i_{id}}$, $\mathbf{e}_{i_{t}}$  and $\mathbf{e}_{i_{v}}$  respectively, process on the target modality $m \in \{id, t, v\}$ as follows:
\begin{equation}
\label{eq:MoE}
    \mathbf{\hat{e}}_{i_m} = \sum_{k=1}^{K} g_{k}^{m} \operatorname{Expert}_{k}(\operatorname{Con(\mathbf{e}_{i_{id}},\mathbf{e}_{i_{t}},\mathbf{e}_{i_{v}})}),
\end{equation}
where $\operatorname{Con(\cdot,\cdot)}$ represents concatenation operation and $\operatorname{Expert}_{*}(\cdot)$ represents the experts. $K$ is a hyper-parameter that denotes the total number of experts for each modality, which we empirically analyzed in the \textbf{Ablation Study} subsection. We adopt feed-forward networks (FFNs) as experts, formally $\operatorname{Expert}_{*}(\cdot)=\operatorname{FFN}(x)=(\operatorname{GeLU}(x \mathbf{W}_1+\mathbf{b}_1)) \mathbf{W}_2+\mathbf{b}_2$, where $\mathbf{W}_1 \in \mathbb{R}^{3d \times 4d}$, $\mathbf{W}_2 \in \mathbb{R}^{4d \times d}$, $\mathbf{b}_1 \in \mathbb{R}^{4d}$, and $\mathbf{b}_2 \in \mathbb{R}^{d}$ are learnable parameters. $g_{k}^{m}$ denotes the corresponding combination weight of the $k$-th expert on modality $m$ from the routing vector $G^m \in \mathbb{R}^{K}$ by attentive gating router:
\begin{equation}
\label{eq:6}
G^m = \operatorname{Softmax}(Q^{m}(K^{m})^T / \sqrt{K}) V^{m},
\end{equation}
\begin{equation}
\label{eq:7}
    Q^{m},K^{m},V^{m} = \mathbf{W}^{m}_{Q} \mathbf{E}_m,\mathbf{W}^{m}_{K} \mathbf{E}_m,\mathbf{W}^{m}_{V} \mathbf{E}_m,
\end{equation}
where $\mathbf{W}^{m}_{Q} \in \mathbb{R}^{d \times K}$, $\mathbf{W}^{m}_{K} \in \mathbb{R}^{d \times K}$, and $\mathbf{W}^{m}_{V} \in \mathbb{R}^{d \times K}$ are learnable parameters. 

Due to the massive amount of user data and items, it is difficult for users to interact with all items, so the recommendation system generally faces the data sparsity problem \cite{xu2025survey,singh2020scalability}, which leads to the challenge of relying on training learnable weights of different modalities. Inspired by previous studies \cite{wang2021personalized,wang2023setrank} leveraging item category information to enhance the understanding of item features, we utilize item categories to guide the allocation of modality weights. Specifically, we treat item categories as true classification labels and compute modality weights by measuring the difference between true classification labels and the predicted labels derived from the corresponding modality information. To generate signals more related to user sequence, we measure the difference between true classification labels and predicted labels by binary cross-entropy loss \cite{du2023ensemble,xie2022decoupled} at the sequence level. For the item $i \in S$ on modality $m$, we employ the representations $\mathbf{e}_{i_m}$ of $i$ to calculate its probability to belong to each class, which is formulated as:
\begin{equation}
\label{eq:8}
    \hat{y}_{c}^{m} = \mathbf{W}_{c}^{m} \mathbf{e}_{i_m} + \mathbf{b}_{c}^{m},
\end{equation}
where $\mathcal{C}$ denotes the set of the categories. $\mathbf{W}_c$ and $\mathbf{b}_c$ are the learnable parameters. $\hat{y}_{c}^{m}$ denote the
likelihood that $c$ belongs to $c \in \mathcal{C}$. Therefore, the difference between true classification labels and predicted labels by binary cross-entropy loss $d^{m}_{i}$ is formulated as:
\begin{equation}
\label{eq:9}
    d^{m}_{i} = -\sum_{c \in \mathcal{C}} y_{c} \log(\hat{y}_{c}^{m}) + (1 - y_{c}) \log(1 - \hat{y}_{c}^{m}),
\end{equation}
where $y_{c}$ is the ground truth\footnote{The categories are accessible to all baselines. All baselines concatenate title, category, and brand fields as text information.}. Then, we further calculate modality weights, formally:
\begin{equation}
\label{eq:fusion}
    w^{m}_i = \frac{\exp({d_{i}^{m}})}{\sum_{n\in \{id, t, v\}}\exp({d_{i}^{n})}}.
\end{equation}

After obtaining modality weights $w^{m}_i$, experts' output $\mathbf{\hat{e}}_{i_{m}}$ and item representation $\mathbf{e}_{i_{m}}$ for each item $i$, we get the final representation $\mathbf{\bar{e}}_{i_{m}}$ for item $i$ on modality $m$, formally:
\begin{equation}
\label{eq:11}
    \mathbf{\bar{e}}_{i_{m}} = \mathbf{e}_{i_{m}} + w^{m}_i \mathbf{\hat{e}}_{i_{m}}.
\end{equation}

\subsection{User Interest Learning}
To effectively capture the dynamic and evolving nature of user interests in sequential interactions, we model user behavior sequences using a Transformer-based encoder \cite{vaswani2017attention}, which has demonstrated strong capabilities in capturing long-range dependencies and temporal patterns in sequential recommendation tasks \cite{kang2018self,sun2019bert4rec}. Specifically, for each modality $m$, we process the enriched item sequential representations $\bar{S}^m = \{\mathbf{\bar{e}}_{i_{m,1}},\mathbf{\bar{e}}_{i_{m,2}},...,\mathbf{\bar{e}}_{i_{m,|S|}}\}$—outputs from our Category-Guided Attentive Mixture of Experts (CAMoE)—through a modality-specific Transformer model. The encoding process is formally defined as:
\begin{equation}
\label{eq:user-interest}
    \mathbf{Z}^m = \operatorname{Transformer}(\operatorname{Dropout}(\operatorname{LayerNorm}(\bar{S}^m))),
\end{equation}
where $\operatorname{Transformer}(\cdot)$ represents Transformer model, $\operatorname{Dropout}(\cdot)$ represents random dropout operation, and $\operatorname{LayerNorm(\cdot)}$ represents layer norm operation. $\mathbf{Z}^{m}$ denotes the final representation corresponding to the last position as user interest under modality $m$. 

Motivated by recent findings \cite{zhu2025transformers} that highlight the role of $\operatorname{LayerNorm(\cdot)}$ in non-linearly scaling inputs and compressing outliers via an S-shaped function akin to $\operatorname{Tanh(\cdot)}$, we explore a lightweight and adaptive alternative: the Dynamic Tanh (DyT) operation, defined as:
\begin{equation}
\label{eq:13}
\operatorname{DyT}(x)=\operatorname{Tanh}(\alpha x),
\end{equation}
where $\alpha$ is a learnable parameter. DyT mimics the normalization and activation behavior of LayerNorm while offering greater efficiency and flexibility by adaptively scaling inputs and constraining output magnitudes via the bounded tanh function. Inspired by its potential to enhance representation quality and training stability, we replace $\operatorname{LayerNorm(\cdot)}$ in Eq.~\ref{eq:user-interest} $\operatorname{DyT}(\cdot)$, leading to:
\begin{equation}
\label{eq:DyT}
    \mathbf{Z}^m = \operatorname{Transformer}(\operatorname{Dropout}(\operatorname{DyT}(\bar{S}^m))).
\end{equation}

In the \textbf{Ablation Study} subsection, we empirically discuss the performance benefits of substituting $\operatorname{LayerNorm(\cdot)}$ with $\operatorname{DyT}(\cdot)$. To predict the interaction probability between user $u$ and item $i$, we first compute a score for each modality and then aggregate these scores by summing them to obtain the final prediction $\hat{y}_i$, formally:
\begin{equation}
\label{eq:15}
    \hat{y}_i = \sum_{m \in \{id,t,v\}} \mathbf{Z}^m \mathbf{e}_{i_{m}}.
\end{equation}

\subsection{Modality Swap Contrastive Learning}
To enhance the learning of correlations between multimodal information, we propose a modality swap contrastive learning task to strengthen the alignment between representations of different modalities. Inspired by a recent study \cite{dufumier2024align}, we construct augmented contrastive views using the modality swap operation. Specifically, for each modality pair $\langle m_1,m_2 \rangle$, where $m_1 \neq m_2$ and $m_1,m_2 \in \{id,t,v\}$, we create a pair of modality swap contrastive views, $\langle \mathbf{Z}_{aug}^{m_1},\mathbf{Z}_{aug}^{m_2} \rangle$, by randomly swapping half of the sequence items by randomly swapping half of the sequence items $\bar{S}^{m_1/m_2} = \{\mathbf{\bar{e}}_{i_{m_1/m_2,1}},...,\mathbf{\bar{e}}_{i_{m_1/m_2,|S|}}\}$ between the two modalities:
\begin{equation}
\label{eq:16}
    \mathbf{Z}_{aug}^{m_1/m_2} = \operatorname{Transformer}(\operatorname{Dropout}(\operatorname{DyT}(\bar{S}^{m_1/m_2}_{aug}))), 
\end{equation}
\begin{equation}
\label{eq:17}
    \bar{S}^{m_1}_{aug}, \bar{S}^{m_2}_{aug} = \operatorname{Swap}( \bar{S}^{m_1}, \bar{S}^{m_2}|0.5),
\end{equation}
where $\operatorname{Swap}(S_1,S_2|\rho)$ denotes randomly swapping the item representations between sequences $S_1$ and $S_2$ with a probability of $\rho$. To better align representations of different modalities, we adopt InfoNCE \cite{oord2018representation} between augmented contrastive views. Specifically, within a training batch $\mathcal{B}$, we can obtain $\{\mathbf{Z}_{aug,1}^{m_1/m_2},\mathbf{Z}_{aug,2}^{m_1/m_2},...,\mathbf{Z}_{aug,|\mathcal{B}|}^{m_1/m_2}\}$, and we treat $\langle \mathbf{Z}_{aug,i}^{m_1},\mathbf{Z}_{aug,i}^{m_2} \rangle$ as the positive pair, while $\{\langle \mathbf{Z}_{aug,i}^{m_1},\mathbf{Z}_{aug,j}^{m_2} \rangle| i\neq j\}$ are regarded as negative pairs. The contrastive learning loss can be formulated as:
\begin{equation}
\label{eq:18}
\mathcal{L}_{MSCL}^{m_{1,2}}= -\sum_{i=1}^{|\mathcal{B}|}\log \frac{\exp (\operatorname{Sim}(\mathbf{Z}_{aug,i}^{m_1}, \mathbf{Z}_{aug,i}^{m_2}))}{\sum_{j=1}^{|\mathcal{B}|} \exp (\operatorname{Sim}(\mathbf{Z}_{aug,i}^{m_1}, \mathbf{Z}_{aug,j}^{m_2}))},
\end{equation}
where $\operatorname{Sim}(\cdot)$ represents cosine similarity. In the \textbf{Ablation Study} section, we conduct an empirical analysis to evaluate the effectiveness of different modality pair(s) choices.

\subsection{Optimization}
We optimize our model using two objectives: modality swap contrastive learning and the main sequential recommendation task. Given a user behavior sequence $S$, the primary goal of sequential recommendation is to predict which item the user is likely to engage with at the $(|S|$$+$$1)$-th time step. Using our proposed model, we obtain the prediction score $\hat{y}_i$, which represents the probability of the user interacting with item $i$ at the next time step. The cross-entropy loss function is then employed as the main training objective to quantify the difference between the predicted score $\hat{y}_i$ and the ground truth ${y}_i$, defined as:
\begin{equation}
\label{eq:19}
    \mathcal{L}_{SR} = - \sum_{i \in \mathcal{I}} {y}_i \log(\hat{y}_i).
\end{equation}

The final learning objective is defined as:
\begin{equation}
\label{eq:20}
    \mathcal{L} = \mathcal{L}_{SR} + \lambda \mathcal{L}_{MSCL},
\end{equation}
where $\lambda$ is the weight hyperparameter. 

To provide a clearer overview of our CAMMSR, we summarize the learning process of CAMMSR in Algorithm~\ref{al}.

\begin{algorithm}[h]
\caption{Procedure of CAMMSR}
\label{al}
\begin{algorithmic} [1] 
\STATE \textbf{Input:} User set $\mathcal{U}$, item set $\mathcal{I}$, modalities $\mathcal{M} = \{id,t,v\}$.
% \STATE \textbf{Output:} Optimization loss $\mathcal{L}$.
\STATE Initialize item ID representation $I_{id}$, item textual representation $f_{i_t}$, item visual representation $f_{i_v}$, and entire representations for each modality $m$ as $\mathbf{E}_m$;
\WHILE{not converged}
\STATE Get representation $\hat{\mathbf{e}}_{i_m}$ for each modality $m$ with attentive mixture of experts via Eq.~\ref{eq:MoE}-\ref{eq:7}.
\STATE Calculate category-guided weight $w_i^m$ for item $i$ on modality $m$ via Eq.~\ref{eq:8}-\ref{eq:fusion}.
\STATE Get final item representation $\mathbf{\bar{e}}_{i_{m}}$ for item $i$ on modality $m$ via Eq.~\ref{eq:11}.
\STATE Get sequences representation $\mathbf{Z}^m$ via Eq.~\ref{eq:13}-\ref{eq:DyT}.
\STATE Compute modality swap contrastive learning loss $\mathcal{L}_{MSCL}$ via Eq.~\ref{eq:16}-\ref{eq:18}.
\STATE Compute main sequential recommendation loss $\mathcal{L}_{SR}$ via Eq.~\ref{eq:15}-\ref{eq:19}.
\STATE Get final optimization loss $\mathcal{L}$ via Eq.~\ref{eq:20}.
\STATE Optimize model via loss $\mathcal{L}$.
\ENDWHILE
\end{algorithmic}
\end{algorithm}
\section{Experiments}
In this section, we conduct comprehensive experiments to evaluate the performance of CAMMSR framework on four widely used real-world datasets. The following five research questions can be well answered through experimental results: \textbf{RQ1: }Does CAMMSR outperform the state-of-the-art conventional and multimodal sequential recommendations? \textbf{RQ2: }What impact do the key components of CAMMSR have on its overall performance? \textbf{RQ3: }How does the category noise affect the performance of CAMMSR? \textbf{RQ4: }Whether leveraging multiple modalities contributes to more ideal performance for CAMMSR? \textbf{RQ5: }How efficient is CAMMSR compared with other multimodal sequential recommendations? \textbf{RQ6: }How does category-guided of CAMMSR assign modality weights effectively? \textbf{RQ7: }How do different hyper-parameter settings impact the overall performance of CAMMSR?

In summary, we introduce the experimental settings in Section~\ref{sec:settings}, demonstrate the performance advantages of CAMMSR in Section~\ref{sec:performance}, evaluate the necessity of key components in Section~\ref{sec:ablation}, analyze CAMMSR in category noise scenarios in Section~\ref{sec:noise}, discuss the importance of multiple modalities in Section~\ref{sec:multimodal}, analyze the efficiency of CAMMSR in Section~\ref{sec:efficiency}, provide a real-world case study in Section~\ref{sec:case}, conduct a hyper-parameter analysis in Section~\ref{sec:hyper-parameter}, and extend CAMMSR to different category label scenarios in Section~\ref{sec:category}.
% \begin{itemize}[leftmargin=*]
%     \item \textbf{RQ1: }Does CAMMSR outperform the state-of-the-art conventional and multimodal sequential recommendations?
%     \item \textbf{RQ2: }What impact do the key components of our CAMMSR have on its overall performance?
%     \item \textbf{RQ3: }How does the category noise affect the performance of our CAMMSR?
%     \item \textbf{RQ4: }Whether leveraging multiple modalities contributes to more ideal performance for our CAMMSR?
%     \item \textbf{RQ5: }How efficient is our CAMMSR compared with other multimodal sequential recommendations?
%     \item \textbf{RQ6: }How does category-guided of CAMMSR assign modality weights effectively?
%     \item \textbf{RQ7: }How do different hyper-parameter settings impact the overall performance of our CAMMSR?
    
% \end{itemize}

\begin{table}[!t]
    \centering
     \caption{Statistics of all four datasets.}
    \label{tab:dataset}
    \vskip -0.05in
    % \small
\resizebox{\linewidth}{!}{
    \begin{tabular}{c|cccc}
     \toprule
         Dataset&  Toys&  Games&  Beauty&  Home\\
         \midrule
         \# Users & 19,412 & 24,303 & 22,363 & 66,520 \\
         \# Items & 11,924 & 10,672 & 12,101 & 28,238 \\
         \# Interactions & 167,597 & 231,780 & 198,502 & 551,682 \\
         \# Avg. Interactions/User & 14.58 & 9.53 & 8.88 & 8.29 \\
         \# Avg. Interactions/Item & 17.03 & 21.72 & 16.40 & 19.54 \\
         \# Sparsity & 99.93\% & 99.91\% & 99.93\% & 99.97\% \\
        \bottomrule
    \end{tabular}
    }

    \vskip -0.1in
\end{table}

\subsection{Experimental Settings}
\label{sec:settings}
\subsubsection{Datasets}
To validate the effectiveness of CAMMSR, we conduct experiments on four public and real-world datasets \cite{he2016ups}, namely \textbf{Toys and Games (Toys)}, \textbf{Video Games (Games)}, \textbf{Beauty}, and \textbf{Home and Kitchen (Home)}. For each dataset, duplicate interactions are removed, and user interactions are sorted chronologically by timestamps to construct behavior sequences. Following prior studies \cite{xu2024aligngroup,hou2022towards}, we apply a 5-core filtering strategy, which excludes users and items with fewer than five interactions from each dataset. For textual data, we concatenate phrases from the title, category, and brand fields, consistent with previous studies \cite{wang2023missrec,xie2022decoupled}. For image data, we directly download the first image of each item from the provided product URLs. Table~\ref{tab:dataset} presents detailed statistical information for all four datasets.

\subsubsection{Baselines}
% To evaluate the effectiveness of our CAMMSR, we compare CAMMSR with various state-of-the-art baselines with two categories: (1) For traditional non-time-aware methods, GRU4Rec \cite{hidasi2015session}, SASRec \cite{kang2018self}, and LRURec \cite{yue2024linear} are included. (2) HM4SR is also compared with multi-modal methods, including NOVA \cite{liu2021noninvasive}, DIFSR \cite{xie2022decoupled}, UniSRec \cite{hou2022towards}, MISSRec \cite{wang2023missrec}, M3SRec \cite{bian2023multi}, IISAN \cite{fu2024iisan}, and TedRec \cite{xu2024sequence}. \red{TODO}
To evaluate the effectiveness of CAMMSR, we compare CAMMSR with various state-of-the-art baselines: 
% \textbf{Sequential Methods} (\textbf{GRU4Rec} \cite{hidasi2015session}, \textbf{SASRec} \cite{kang2018self}, \textbf{LRURec} \cite{yue2024linear}, \textbf{FEARec} \cite{du2023frequency}, and \textbf{TiCoSeRec} \cite{dang2023ticoserec}) and \textbf{Multimodal Sequential Methods} (\textbf{NOVA} \cite{liu2021noninvasive}, \textbf{DIF-SR} \cite{xie2022decoupled}, \textbf{UniSRec} \cite{hou2022towards}, \textbf{MISSRec} \cite{wang2023missrec}, \textbf{M3SRec} \cite{bian2023multi}, \textbf{IISAN} \cite{fu2024iisan}, \textbf{MMMLP} \cite{liang2023mmmlp}, and \textbf{TedRec} \cite{xu2024sequence}). Details for all baselines can be found in the \textbf{Supplementary Material}.

\noindent \textbf{(1) Sequential Methods} 
\begin{itemize}[leftmargin=*]
    \item \textbf{GRU4Rec} \cite{hidasi2015session} utilizes Gated Recurrent Units (GRU) to model user click sequences for the session-based recommendation. Following previous studies \cite{bian2023multi,sun2019bert4rec}, we represent the items using embedding vectors rather than one-hot vectors.
    \item \textbf{SASRec} \cite{kang2018self} is a sequential recommendation model based on self-attention, leveraging the multi-head attention mechanism to predict the next item.
    \item \textbf{LRURec} \cite{yue2024linear} introduces linear recurrent units to enhance the encoding quality of long user behavior sequences.
    \item \textbf{FEARec} \cite{du2023frequency} performs frequency-level sequence analysis by introducing a ramp structure to process information from the frequency domain effectively.
    \item \textbf{TiCoSeRec} \cite{dang2023ticoserec} proposes five-time interval-based sequence-level augmentations to create uniform sequences and leverages contrastive learning to enhance performance.
\end{itemize}

\noindent \textbf{(2) Multimodal Sequential Methods}
\begin{itemize}[leftmargin=*]
    \item \textbf{NOVA} \cite{liu2021noninvasive} proposes a non-invasive self-attention mechanism to leverage side information effectively.
    \item \textbf{DIF-SR} \cite{xie2022decoupled} introduces a decoupled side information fusion module to address the rank bottleneck and enhance the expressiveness of non-invasive attention matrices.
    \item \textbf{UniSRec} \cite{hou2022towards} uses item textual information to learn more transferable representations for sequential recommendation.
    \item \textbf{MISSRec} \cite{wang2023missrec} proposes an Interest Discovery Module to capture deep relationships among items and user preferences.
    \item \textbf{M3SRec} \cite{bian2023multi} employs modal-specific and cross-modal MoE mechanisms to enhance modality learning in Transformers.
    \item \textbf{IISAN} \cite{fu2024iisan} utilizes decoupled parameter-efficient fine-tuning on modality foundation models to enable both intra-modal and inter-modal adaptation.
    \item \textbf{MMMLP} \cite{liang2023mmmlp} proposes a purely MLP-based model with Feature Mixer and Fusion Mixer to process multimodal data, achieving outstanding performance with linear complexity.
    \item \textbf{TedRec} \cite{xu2024sequence} performs a sequence-level semantic fusion of text and ID using FFT in the frequency domain.
    \item \textbf{FindRec} \cite{wang2025findrec} theoretically guarantees cross-modal consistency through the Stein kernel-based Integrated Information Coordination Module. 
\end{itemize}

\begin{table*}[!t]
 \vskip -0.1in
\caption{Performance comparison of baselines on different datasets in terms of NDCG@K and MRR@K. The superscript $^*$ indicates the improvement is statistically significant where the $p$-value is less than 0.05. ``Neg. Ratio" denotes the proportion of the test set where removing CG leads to improved recommendation accuracy.}
\label{tab:comparison results}
 \vskip -0.05in
\centering
\setlength{\tabcolsep}{1.55mm}
\resizebox{\linewidth}{!}{
    \begin{tabular}{ccccclcccclcccclcccc}
     \toprule
         \multirow{4}{*}{Baseline} & \multicolumn{4}{c}{Toys}&& \multicolumn{4}{c}{Games}&&  \multicolumn{4}{c}{Beauty}&&  \multicolumn{4}{c}{Home}\\\cmidrule{2-5} \cmidrule{7-10} \cmidrule{12-15} \cmidrule{17-20} &\multicolumn{2}{c}{NDCG} & \multicolumn{2}{c}{MRR} && \multicolumn{2}{c}{NDCG} & \multicolumn{2}{c}{MRR} && \multicolumn{2}{c}{NDCG} & \multicolumn{2}{c}{MRR} && \multicolumn{2}{c}{NDCG} & \multicolumn{2}{c}{MRR}\\ \cmidrule{2-5} \cmidrule{7-10} \cmidrule{12-15} \cmidrule{17-20}
         & @5& @10& @5& @10&&@5& @10& @5& @10&&@5& @10& @5& @10&&@5& @10& @5& @10\\\midrule
         GRU4Rec & 0.0236 & 0.0289 & 0.0201 & 0.0222 && 0.0385 & 0.0514 & 0.0311 & 0.0365 && 0.0271 & 0.0337 & 0.0223 & 0.0260 && 0.0067 & 0.0087 & 0.0056 & 0.0064 \\ 
         SASRec & 0.0348& 0.0411& 0.0257& 0.0295&& 0.0391& 0.0546& 0.0286& 0.0349&& 0.0325& 0.0412& 0.0250& 0.0286&& 0.0118& 0.0148& 0.0089& 0.0100\\ 
         LRURec & 0.0362& 0.0446& 0.0278& 0.0312&& 0.0410& 0.0557& 0.0315& 0.0376&& 0.0323& 0.0412& 0.0250& 0.0286&& 0.0112& 0.0141& 0.0084& 0.0096\\ 
         FEARec & 0.0354& 0.0442& 0.0266& 0.0302&& 0.0411& 0.0558& 0.0306& 0.0367&& 0.0342& 0.0434& 0.0265& 0.0304&& 0.0121& 0.0152& 0.0092& 0.0105\\
         TiCoSeRec & 0.0336& 0.0404& 0.0280& 0.0311&& 0.0463& 0.0589& 0.0382& 0.0434&& 0.0335& 0.0413& 0.0285& 0.0317&& 0.0108& 0.0134& 0.0091& 0.0102\\\midrule
         NOVA& 0.0381& 0.0478& 0.0288& 0.0328&& 0.0417& 0.0576& 0.0303& 0.0368&& 0.0347& 0.0442& 0.0273& 0.0312&& 0.0120& 0.0147& 0.0092& 0.0102\\
         DIF-SR& 0.0359& 0.0444& 0.0284& 0.0318&& 0.0416& 0.0575& 0.0311& 0.0375&& 0.0340& 0.0436& 0.0262& 0.0302&& 0.0092& 0.0112& 0.0078& 0.0086\\
         UniSRec& 0.0276& 0.0384& 0.0194& 0.0239&& 0.0386& 0.0535& 0.0291& 0.0353&& 0.0287& 0.0393& 0.0212& 0.0288&& 0.0121& 0.0160& 0.0092& 0.0108\\
         MISSRec& 0.0323& 0.0414& 0.0235& 0.0270&& 0.0398& 0.0506& 0.0312& 0.0353&& 0.0315& 0.0397& 0.0240& 0.0271&& 0.0140& 0.0174& 0.0107& 0.0119\\
         M3SRec& \underline{0.0416}& 0.0486& \underline{0.0363}& 0.0391&& 0.0493& 0.0643& 0.0404& 0.0464&& 0.0345& 0.0428& 0.0294& 0.0328&& 0.0122& 0.0144& 0.0105& 0.0115\\
         IISAN& 0.0395& \underline{0.0489}& 0.0354& \underline{0.0393}&& \underline{0.0525}& \underline{0.0671}& \underline{0.0432}& \underline{0.0491}&& \underline{0.0372}& \underline{0.0464}& \underline{0.0309}& \underline{0.0347}&& \underline{0.0152}& \underline{0.0187}& \underline{0.0129}& \underline{0.0142}\\
         MMMLP& 0.0378& 0.0466& 0.0344& 0.0377&& 0.0486& 0.0644& 0.0397& 0.0464&& 0.0351& 0.0433& 0.0295& 0.0333&& 0.0133& 0.0165& 0.0107& 0.0118\\
         TedRec& 0.0318& 0.0397& 0.0265& 0.0297&& 0.0468& 0.0604& 0.0384& 0.0439&& 0.0330& 0.0419& 0.0275& 0.0311&& 0.0113& 0.0140& 0.0094& 0.0105\\
         FindRec& 0.0390& 0.0480& 0.0357& 0.0389&& 0.0501& 0.0663& 0.0411& 0.0477&& 0.0359& 0.0447& 0.0304& 0.0341&& 0.0137& 0.0171& 0.0111& 0.0124\\\midrule
         CAMMSR& \textbf{0.0477$^*$}& \textbf{0.0542$^*$}& \textbf{0.0421$^*$}& \textbf{0.0449$^*$}&& \textbf{0.0560$^*$}& \textbf{0.0701$^*$}& \textbf{0.0474$^*$}& \textbf{0.0533$^*$}&& \textbf{0.0429$^*$}& \textbf{0.0517$^*$}& \textbf{0.0374$^*$}& \textbf{0.0404$^*$}&& \textbf{0.0168$^*$}& \textbf{0.0199$^*$}& \textbf{0.0149$^*$}& \textbf{0.0159$^*$}\\
         Improv. & 14.66\%& 10.84\%& 15.98\%& 14.25\%&& 6.67\%& 4.47\%& 9.72\%& 8.55\%&& 15.32\%& 11.42\%& 21.04\%& 16.43\%&& 10.53\%& 6.42\%& 15.50\%& 11.97\%\\
         Neg. Ratio & 0.16\%& 0.16\%& 0.24\%& 0.24\%&& 0.33\%& 0.33\%& 0.61\%& 0.61\%&& 0.24\%& 0.24\%& 0.40\%& 0.40\%&& 0.28\%& 0.28\%& 0.54\%& 0.54\%\\
         \bottomrule
    \end{tabular}
    }
    \vskip -0.1in
\end{table*}

\begin{table}[t]
\centering
 \vskip -0.05in
\caption{Effectiveness of CAMMSR with different modality pair(s) choices. \faCircle $ $ denotes chosen pair(s).}
\label{tab:ablation2}
 \vskip -0.05in
\resizebox{\linewidth}{!}{
    \begin{tabular}{c|ccc|cccc}
    \toprule
        Dataset & $\langle v, t \rangle$ & $\langle id, t \rangle$ & $\langle id, v \rangle$ & NDCG@5 & NDCG@10 & MRR@5 & MRR@10\\ \cmidrule{1-8}
        \multirow{8}{*}{Toys} & \faCircleO & \faCircleO & \faCircleO & 0.0415& 0.0504& 0.0371& 0.0415\\
        & \faCircle & \faCircleO & \faCircleO & 0.0435& 0.0530& 0.0396& 0.0439\\
        & \faCircleO & \faCircle & \faCircleO & 0.0424& 0.0516& 0.0382& 0.0422\\
        & \faCircleO & \faCircleO & \faCircle & 0.0421& 0.0513& 0.0380& 0.0421\\
        & \faCircle & \faCircle & \faCircleO & 0.0439& 0.0536& 0.0406& 0.0444\\
        & \faCircle & \faCircleO & \faCircle & 0.0439& 0.0538& 0.0407& 0.0446\\
        & \faCircleO & \faCircle & \faCircle & 0.0430& 0.0524& 0.0387& 0.0427\\
        & \faCircle & \faCircle & \faCircle & \textbf{0.0447}& \textbf{0.0542}& \textbf{0.0421}& \textbf{0.0449}\\\cmidrule{1-8}
        \multirow{8}{*}{Games} & \faCircleO & \faCircleO & \faCircleO & 0.0529& 0.0679& 0.0436& 0.0501\\
        & \faCircle & \faCircleO & \faCircleO & 0.0541& 0.0690& 0.0455& 0.0525\\
        & \faCircleO & \faCircle & \faCircleO & 0.0535& 0.0686& 0.0442& 0.0508\\
        & \faCircleO & \faCircleO & \faCircle & 0.0533& 0.0684& 0.0440& 0.0505\\
        & \faCircle & \faCircle & \faCircleO & 0.0550& 0.0693& 0.0462& 0.0527\\
        & \faCircle & \faCircleO & \faCircle & 0.0552& 0.0696& 0.0465& 0.0529\\
        & \faCircleO & \faCircle & \faCircle & 0.0545& 0.0689& 0.0449& 0.0515\\
        & \faCircle & \faCircle & \faCircle & \textbf{0.0560}& \textbf{0.0701}& \textbf{0.0474}& \textbf{0.0533}\\\cmidrule{1-8}
        \multirow{8}{*}{Beauty} & \faCircleO & \faCircleO & \faCircleO & 0.0374& 0.0469& 0.0312& 0.0350\\
        & \faCircle & \faCircleO & \faCircleO & 0.0414& 0.0503& 0.0367& 0.0392\\
        & \faCircleO & \faCircle & \faCircleO & 0.0383& 0.0479& 0.0330& 0.0365\\
        & \faCircleO & \faCircleO & \faCircle & 0.0381& 0.0479& 0.0328& 0.0363\\
        & \faCircle & \faCircle & \faCircleO & 0.0417& 0.0506& 0.0367& 0.0394\\
        & \faCircle & \faCircleO & \faCircle & 0.0418& 0.0509& 0.0369& 0.0396\\
        & \faCircleO & \faCircle & \faCircle & 0.0388& 0.0492& 0.0343& 0.0378\\
        & \faCircle & \faCircle & \faCircle & \textbf{0.0429}& \textbf{0.0517}& \textbf{0.0374}& \textbf{0.0404}\\\cmidrule{1-8}
        \multirow{8}{*}{Home} & \faCircleO & \faCircleO & \faCircleO & 0.0152& 0.0185& 0.0128& 0.0142\\
        & \faCircle & \faCircleO & \faCircleO & 0.0162& 0.0191& 0.0141& 0.0152\\
        & \faCircleO & \faCircle & \faCircleO & 0.0155& 0.0188& 0.0131& 0.0146\\
        & \faCircleO & \faCircleO & \faCircle & 0.0155& 0.0187& 0.0133& 0.0145\\
        & \faCircle & \faCircle & \faCircleO & 0.0163& 0.0193& 0.0142& 0.0153\\
        & \faCircle & \faCircleO & \faCircle & 0.0165& 0.0195& 0.0147& 0.0156\\
        & \faCircleO & \faCircle & \faCircle & 0.0160& 0.0190& 0.0140& 0.0148\\
        & \faCircle & \faCircle & \faCircle & \textbf{0.0168}& \textbf{0.0199}& \textbf{0.0149}& \textbf{0.0159}\\
        \bottomrule
         
    \end{tabular}
    }
    \vskip -0.2in
  
    % \vskip -0.1in
\end{table}

\subsubsection{Evaluation Metrics}
To fairly evaluate the performance of the recommendation task, we follow previous studies \cite{liang2023mmmlp,wang2023missrec} and utilize two established metrics: Normalized Discounted Cumulative Gain (NDCG@K) and Mean Reciprocal Rank (MRR@K), where K $\in \{5, 10\}$. We adopt the leave-one-out evaluation strategy to conduct the experiments. The ranking scores are computed on the whole item set without sampling.

\subsubsection{Implementation Details}
To ensure a fair comparison, we implement our CAMMSR and all baselines using the RecBole \cite{zhao2021recbole}. RecBole is a unified and comprehensive framework for recommendation systems, while RecBole implements various recommendation methods. We set the training batch size to 1024, and the hidden embedding size for all baselines is 64. The maximum length of each behavior sequence is limited to 50. For the encoder structure, the number of multi-heads in the Transformer and the number of self-attention layers are both empirically set to 2. Specifically, we employ the same Adam \cite{kingma2014adam} optimizer and Xavier initialization \cite{glorot2010understanding} with default parameters. We conduct a detailed hyper-parameter search for all baseline models. The expert number $K$ for our CAMoE component is conducted within the range of $\{1,2,4,8,16\}$ weight hyper-parameter $\lambda$ within the range of $\{0.25,0.5,0.75,1\}$. In addition, we use empirically fixed $1e^{-3}$ as the learning rate. All experiments are conducted on a single NVIDIA GeForce RTX 3090 GPU.

\begin{figure}[h]
    \centering
     \vskip -0.1in
    \includegraphics[width=1\linewidth]{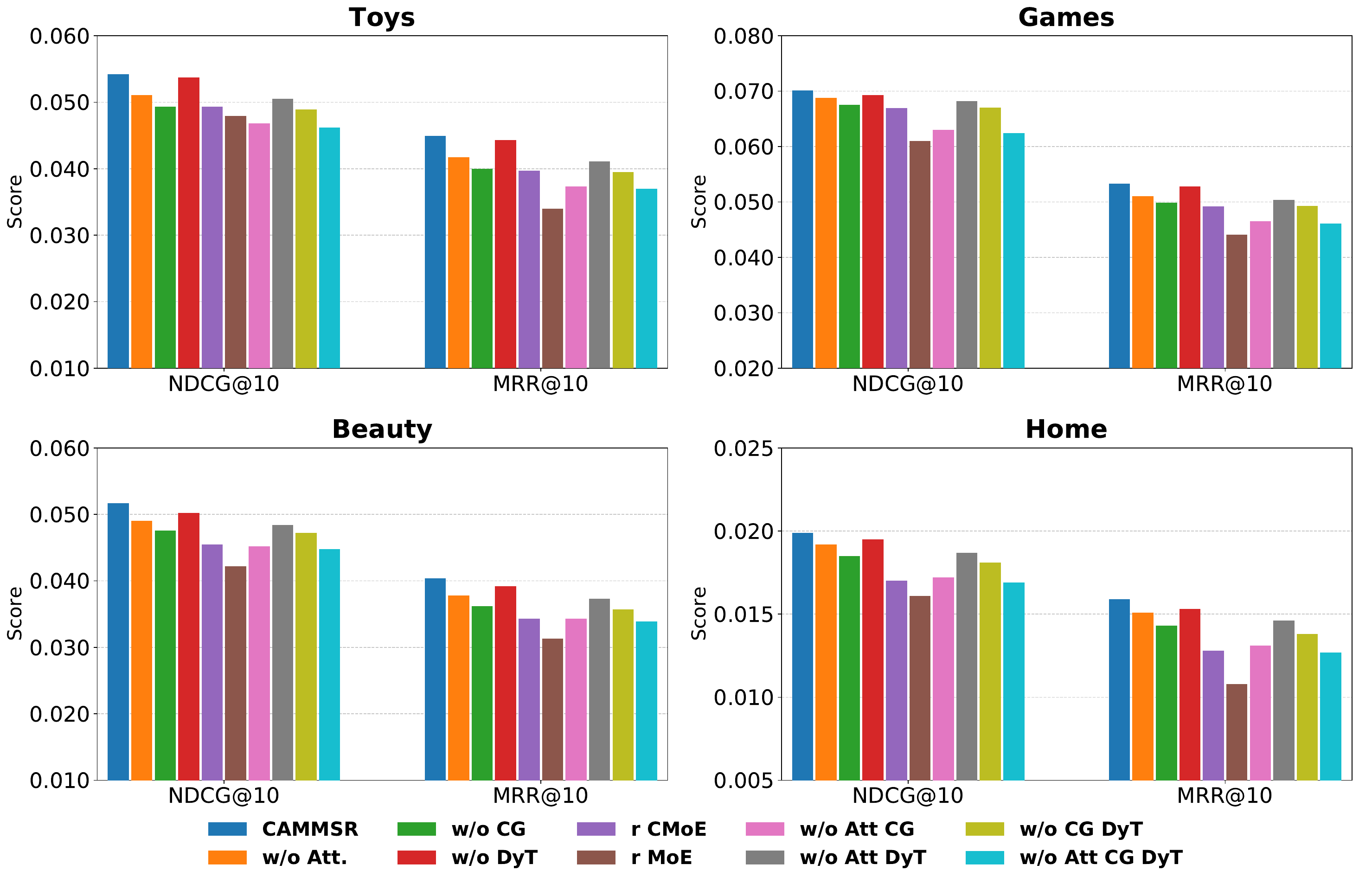}
     \vskip -0.1in
    \caption{Performance comparison for CAMMSR and all variants across all four datasets.}
     \vskip -0.15in
    \label{fig:ablation study}
\end{figure}

\subsection{Overall Comparison (RQ1)}
\label{sec:performance}
Table~\ref{tab:comparison results} presents the comprehensive recommendation performance of all baseline models across four publicly available datasets. The best-performing result is highlighted in bold, while the second-best result is underlined. We also report the relative improvement (\%Improv.) achieved by our proposed CAMMSR, with statistical significance verified through t-test experiments ($p$ $<$ 0.05) as indicated by the superscript $*$. Based on these experimental results, we provide the following detailed observations:

\begin{itemize}[leftmargin=*]
    \item CAMMSR demonstrates consistent and statistically significant superiority across all evaluation metrics and datasets. Specifically, on the Toys dataset, it achieves relative improvements of 14.66\% in NDCG@5 and 15.98\% in MRR@5 compared to the strongest baseline. Similar substantial gains are observed across other datasets, with improvements of 6.67\% in NDCG@5 on Games, 15.32\% in NDCG@5 on Beauty, and 10.53\% in NDCG@5 on Home. This consistent performance advantage validates that our category-guided attentive MoE framework, combined with modality swap contrastive learning, effectively mines discriminative modality features and leverages synergistic effects between modalities for sequential recommendation.
    \item The comparative analysis reveals that multimodal sequential methods generally outperform traditional sequential approaches across most experimental settings. For instance, on the Beauty dataset, all multimodal baselines surpass traditional methods by substantial margins, with the strongest multimodal baseline IISAN achieving 0.0372 in NDCG@5 compared to 0.0342 from the best traditional method FEARec. Notably, CAMMSR further advances this performance frontier, demonstrating the critical importance of our tailored category-guided attentive MoE component in dynamically assigning appropriate weights to different modalities based on contextual signals.
    \item The results highlight that effectively modeling inter-modal relationships is crucial for recommendation performance. While IISAN and M3SRec achieve competitive results through their dedicated inter-modal modeling approaches (e.g., IISAN reaches 0.0525 in NDCG@5 on Games), CAMMSR consistently surpasses them across all datasets and metrics. This performance advantage is particularly evident in the Beauty dataset, where CAMMSR achieves a 15.32\% improvement in NDCG@5 over IISAN, demonstrating the superior capability of our category-guided attentive MoE in exploiting synergistic effects between different modalities through adaptive expert combination and contrastive alignment.
    \item While it is possible that extreme cases, where category-modality correlation diverges from recommendation utility, might occur by chance, we address this concern empirically by analyzing the proportion of the test set where removing CAMoE leads to improved recommendation accuracy. Experimental results show that CAMoE has a negative effect in less than 1\% of cases. This demonstrates its strong generalizability and stability. From the perspective of performance improvement, CAMMSR shows more significant performance advantages in scenarios with lower functional requirements (e.g., Toys and Beauty) compared to those with higher functional requirements (e.g., Games and Home).
\end{itemize}

Overall, the comprehensive experimental results confirm that CAMMSR delivers statistically significant and practically meaningful improvements in recommendation performance.

\subsection{Ablation Study (RQ2)}
\label{sec:ablation}
To analyze the effectiveness of our CAMMSR, we conduct comprehensive ablation studies to evaluate the necessity and contribution of each individual component within the model. Specifically, we compare our model with the following variants: a) \textbf{$\mathbf{w/o}$ Att.} replaces the attentive gating routers in Eq.~\ref{eq:MoE} with linear gating routers. b) \textbf{$\mathbf{w/o}$ CG} replaces category-guided fusion Eq.~\ref{eq:fusion} with equal weight fusion. c) \textbf{$\mathbf{w/o}$ DyT} replaces the DyT layer in Eq.~\ref{eq:DyT} with the LayerNorm layer. d) \textbf{$\mathbf{r}$ CMoE} replaces our CAMoE with cross-modal MoE from previous work \cite{bian2023multi}. e) \textbf{$\mathbf{r}$ MoE} replaces our CAMoE with vanilla MoE. The evaluation results are demonstrated in Figure~\ref{fig:ablation study}. All the components contribute to the performance of CAMMSR. Specifically, we have the following observations: 1) CAMMSR achieves higher performance than \textbf{$\mathbf{w/o}$ Att.}, demonstrating the necessity of giving appropriate weights to different experts in MoE. 2) The performance of \textbf{$\mathbf{w/o}$ CG} is significantly lower than that of CAMMSR, which demonstrates that assigning weights based on the performance of different modalities in the category-prediction task effectively guides how to leverage multimodal features. 3) CAMMSR achieves notable performance improvements over \textbf{$\mathbf{w/o}$ DyT}, demonstrating that the DyT layer can achieve comparable or even superior performance compared to the LayerNorm layer in Transformers. This finding aligns with the observations reported in previous studies \cite{zhu2025transformers}. 4) CAMMSR consistently outperforms \textbf{$\mathbf{r}$ CMoE} and \textbf{$\mathbf{r}$ MoE}, demonstrating the effectiveness of our CAMoE. Furthermore, we further investigate the impact of different modality pair selections in the Modality Swap Contrastive Learning component on the overall performance of CAMMSR. Table~\ref{tab:ablation2} illustrates the investigation result. We observe that incorporating more modality pairs leads to higher performance. When selecting only a single modality pair, the combination of visual and textual modalities proves to be the optimal choice. We hypothesize that this is due to the significant semantic gap between visual and textual modalities, which the Modality Swap Contrastive Learning component effectively mitigates, thereby enabling better utilization of multimodal features.

% \begin{figure}[ht]
%     \centering
%      % \vskip -0.1in
%     \includegraphics[width=1\linewidth]{pic/Combined.pdf}
%      % \vskip -0.1in
%     \caption{Analysis of removing each modality.}
%      % \vskip -0.2in
%     \label{fig:in-depth}
% \end{figure}

\begin{figure*}
    \centering
     \vskip -0.1in
    \includegraphics[width=0.85\linewidth]{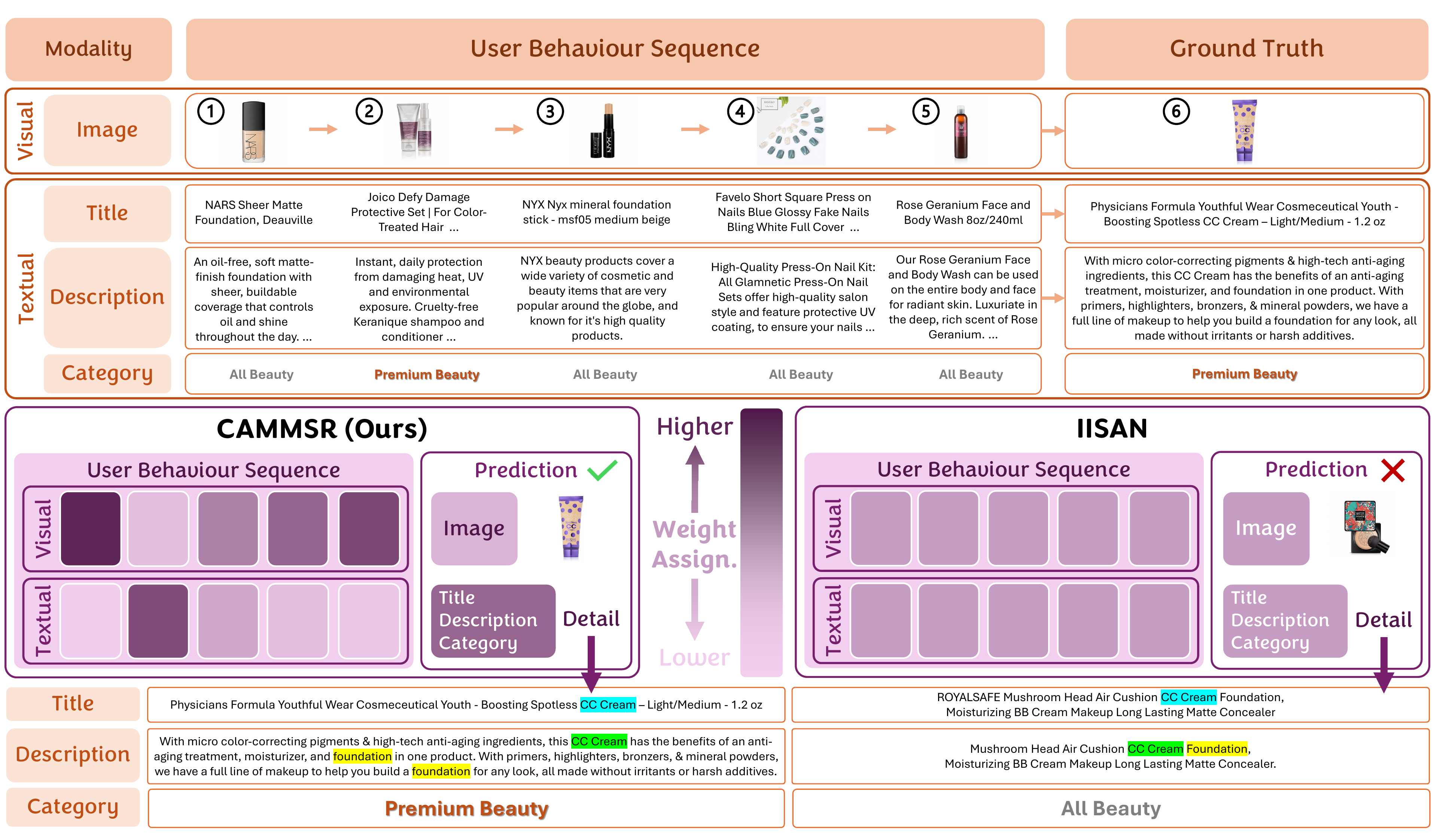}
     \vskip -0.1in
    \caption{A case study for purchase sequence and CAMMSR and IISAN prediction results from the Beauty Dataset.}
     \vskip -0.1in
    \label{fig:case study}
\end{figure*}

\begin{figure}[ht]
    \centering
     \vskip -0.05in
    \subfigure[Toys] {
        \label{fig:toys}
        \includegraphics[width=0.42\linewidth]{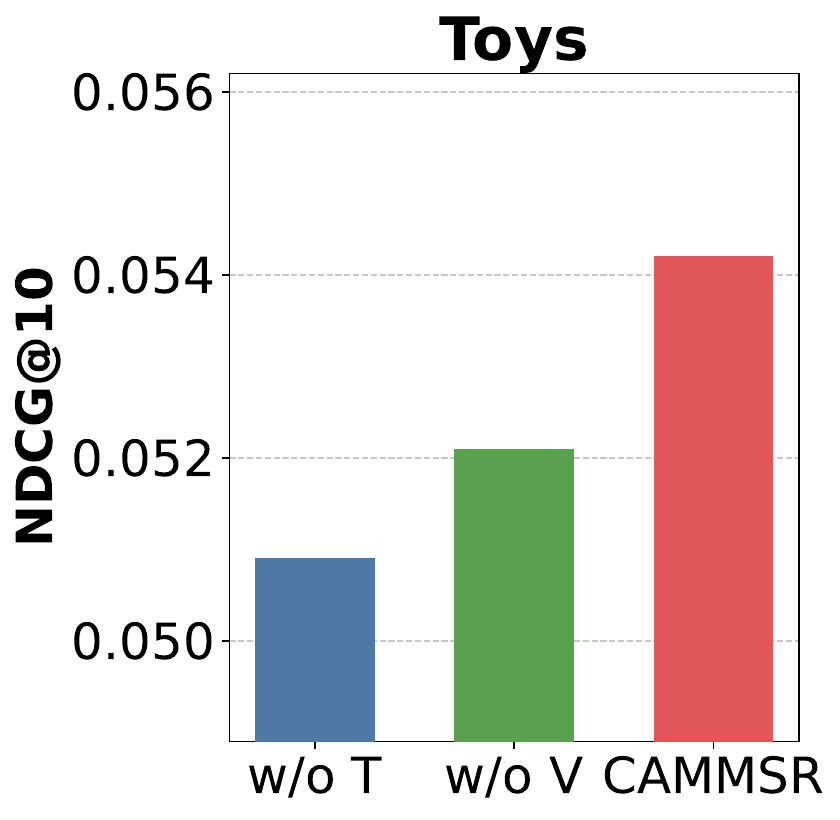}
        }
    \subfigure[Games] {
        \label{fig:games}
        \includegraphics[width=0.42\linewidth]{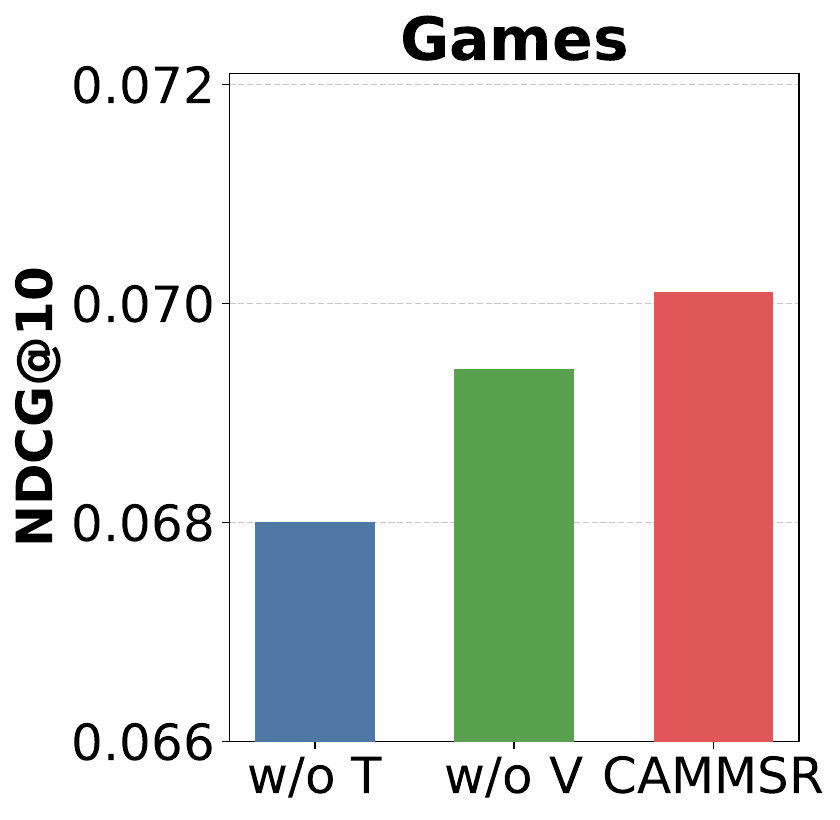}
        }
    \subfigure[Beauty] {
        \label{fig:beauty}
        \includegraphics[width=0.42\linewidth]{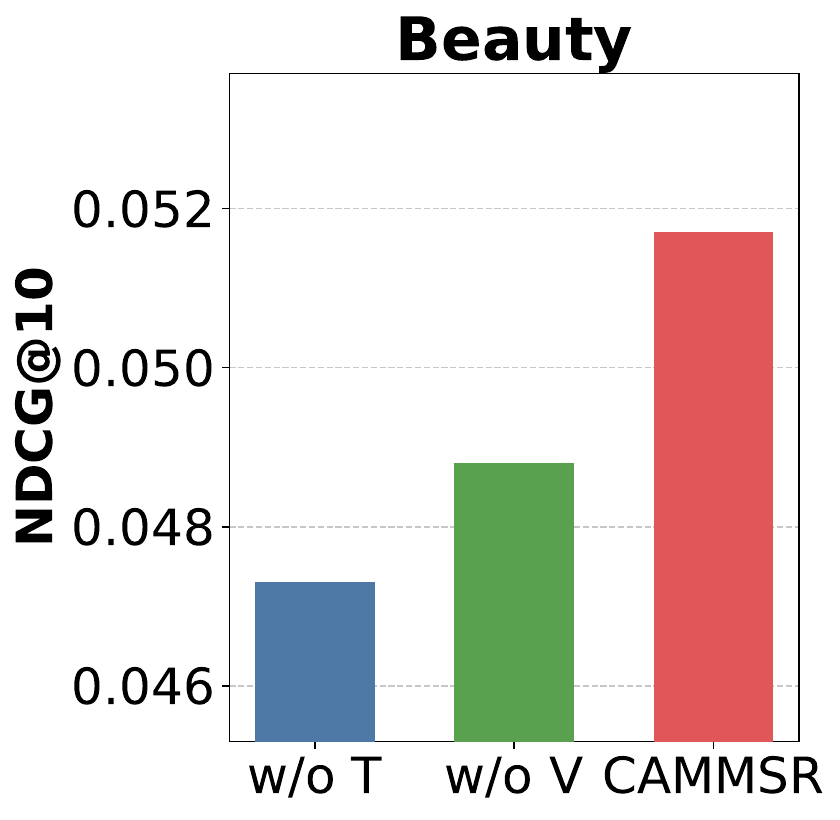}
        }
    \subfigure[Home] {
        \label{fig:home}
        \includegraphics[width=0.42\linewidth]{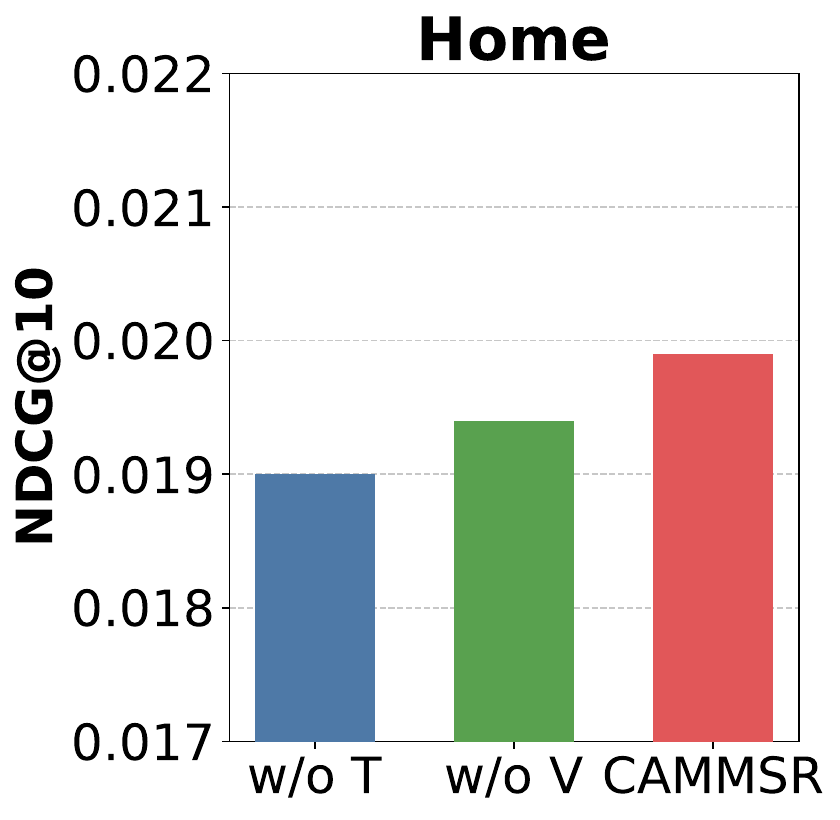}
        }
        \vskip -0.1in
    \caption{Performance comparison of removing each modality.}
    \label{fig:in-depth}
    \vskip -0.2in
\end{figure}

\begin{table}[!t]
    \centering
    \small
    \vskip -0.1in
    \caption{Depth analysis for category noise.}
    \label{tab:noise}
    \vskip -0.05in
    \resizebox{\linewidth}{!}{
    \begin{tabular}{cccclcc}
    \toprule
        \multirow{2}{*}{Type}& \multirow{2}{*}{Level} & \multicolumn{2}{c}{Toys}& & \multicolumn{2}{c}{Games} \\ 
        \cmidrule{3-4} \cmidrule{6-7} 
        & & NDCG@10 & MRR@10 && NDCG@10 & MRR@10 \\ \midrule
        CAMMSR & - & 0.0542 & 0.0449 && 0.0701 & 0.0533  \\ \midrule
        \multirow{3}{*}{Removal} & 10\% & 0.0530 & 0.0439 && 0.0693 & 0.0524 \\ 
        & 20\% & 0.0520 & 0.0428 && 0.0689 & 0.0519  \\ 
        & 30\% & 0.0513 & 0.0420 && 0.0684 & 0.0515  \\ \midrule
        \multirow{3}{*}{Misleading} & 10\% & 0.0525 & 0.0433 && 0.0688 & 0.0519 \\ 
        & 20\% & 0.0505 & 0.0411 && 0.0678 & 0.0504 \\ 
        & 30\% & 0.0482 & 0.0391 && 0.0667 & 0.0490 \\ \midrule
        $w/o$ CG & - & 0.0493 & 0.0400 && 0.0675 & 0.0499 \\ \bottomrule
        \toprule
        \multirow{2}{*}{Type}& \multirow{2}{*}{Level} & \multicolumn{2}{c}{Beauty}& & \multicolumn{2}{c}{Home} \\ 
        \cmidrule{3-4} \cmidrule{6-7}
        && NDCG@10 & MRR@10 && NDCG@10 & MRR@10 \\ \midrule
        CAMMSR & - & 0.0517 & 0.0404 && 0.0199 & 0.0159 \\ \midrule
        \multirow{3}{*}{Removal} & 10\% & 0.0505 & 0.0391 && 0.0195 & 0.0156 \\ 
        & 20\% & 0.0499 & 0.0384 && 0.0193 & 0.0152 \\ 
        & 30\% & 0.0492 & 0.0377 && 0.0190 & 0.0148 \\ \midrule
        \multirow{3}{*}{Misleading} & 10\% & 0.0500 & 0.0386 && 0.0191 & 0.0150 \\ 
        & 20\% & 0.0480 & 0.0368 && 0.0186 & 0.0146 \\ 
        & 30\% & 0.0470 & 0.0357 && 0.0182 & 0.0142 \\ \midrule
        $w/o$ CG & - & 0.0476 & 0.0362 && 0.0185 & 0.0143 \\ \bottomrule
    \end{tabular}
    }
    \vskip -0.1in
    
\end{table}

\subsection{Noise Study (RQ3)}
\label{sec:noise}
To evaluate the robustness of our category-guided mechanism under realistic data quality scenarios, we conduct a systematic noise study simulating two common types of category label corruption: (1) \textbf{Removal}, where corrupted tags are deleted and equal weights are assigned to all modalities as fallback; and (2) \textbf{Misleading}, where corrupted tags are replaced with incorrect labels. For reference, we include the $w/o$ CG variant that completely removes category-guided weight allocation. The hyperparameter configurations remain consistent with our main experiments (Section~\ref{sec:settings}). Experimental results across all four datasets are summarized in Table~\ref{tab:noise}, from which we derive the following key observations:

\begin{itemize}[leftmargin=*]
    \item The \textbf{Removal} condition consistently results in less performance degradation compared to the \textbf{Misleading} scenario across all noise levels. For instance, on the Toys dataset with 30\% noise, Removal causes a 5.35\% drop in NDCG@10, while Misleading leads to an 11.07\% decrease. This pattern highlights the critical importance of accurate category signals for proper modality weight allocation and underscores the vulnerability of multimodal systems to misleading supervisory signals.
    \item Even under substantial noise conditions, our category-guided mechanism maintains its effectiveness. At 20\% misleading noise, CAMMSR still achieves superior performance (NDCG@10: 0.0505 on Toys, 0.0678 on Games) compared to the $w/o$ CG variant (NDCG@10: 0.0493 on Toys, 0.0675 on Games), demonstrating considerable robustness to label inaccuracies. This resilience is particularly valuable for real-world deployment where perfect label quality cannot be guaranteed.
\end{itemize}

Based on these empirical findings, we provide the following analysis regarding practical implications:

\begin{itemize}[leftmargin=*]
    \item In real-world data management scenarios, category label errors exceeding 10-20\% are statistically rare in properly maintained systems. Our results confirm that within this realistic noise range, the category-guided component continues to provide net positive benefits, maintaining performance superior to ablation without category guidance. This makes CAMMSR particularly suitable for practical passive data management systems where perfect label curation is economically infeasible.
    \item The significant performance gap between Removal and Misleading conditions (e.g., 0.0513 vs. 0.0482 in NDCG@10 on Toys with 30\% noise) suggests that developing effective label noise filtering mechanisms represents a promising research direction. Future work could focus on integrating confidence-aware category estimation or self-correcting label refinement to further enhance robustness in noisy environments, thereby advancing the reliability of passive data management systems in web-scale applications.
\end{itemize}

\subsection{In-depth Analysis (RQ4)}
\label{sec:multimodal}
We conduct an in-depth analysis to verify whether leveraging multiple modalities contributes to more accurate recommendations. Specifically, we designed the following two variants: a) \textbf{$\mathbf{w/o}$ T} removes textual modality. b) \textbf{$\mathbf{w/o}$ V} removes visual modality. 

As illustrated in Figure~\ref{fig:in-depth}, CAMMSR demonstrates superior performance compared to all its variants. Additionally, the variant \textbf{$\mathbf{w/o}$ V} outperforms \textbf{$\mathbf{w/o}$ T}, indicating the relative importance of textual data over visual data. This phenomenon can be attributed to the ability of multiple modalities to provide richer and more comprehensive information. Moreover, textual information tends to exhibit a stronger semantic alignment with item characteristics, making it more directly relevant to the recommendation task compared to visual information.

\subsection{Efficiency Study (RQ5)}
\label{sec:efficiency}
To further analyze the training efficiency of CAMMSR, we plot scatter charts of efficiency (training time per epoch) and effectiveness (NDCG@10) on Games and Beauty datasets. As shown in Figure~\ref{fig:efficiency}, we can observe that CAMMSR achieves advanced efficiency and performance, outperforming all baselines in terms of efficiency—with the sole exception of DIF-SR—and significantly surpassing all baselines in performance. Furthermore, considering that the total training time is closely related to the model's convergence speed, we also provide an analysis that takes the number of epochs to converge into account. In Table~\ref{tab:efficiency}, we present the efficiency of CAMMSR and advanced baselines in terms of average training time per epoch, the number of epochs required for convergence, and total training time. The results demonstrate that CAMMSR achieves competitive per-epoch training efficiency and the fastest convergence among all models. This highlights that CAMMSR not only reduces computational time but also reaches optimal performance swiftly, making it particularly suitable for scenarios with limited time or computational resources, such as rapid deployment and frequent updates. Additionally, in terms of inference time, CAMMSR achieves better efficiency than other baselines, second only to DIF-SR.

\begin{figure}[h]
    \centering
    \vskip -0.1in
    \subfigure[Games] {
        \label{fig:m}
        \includegraphics[width=0.8\linewidth]{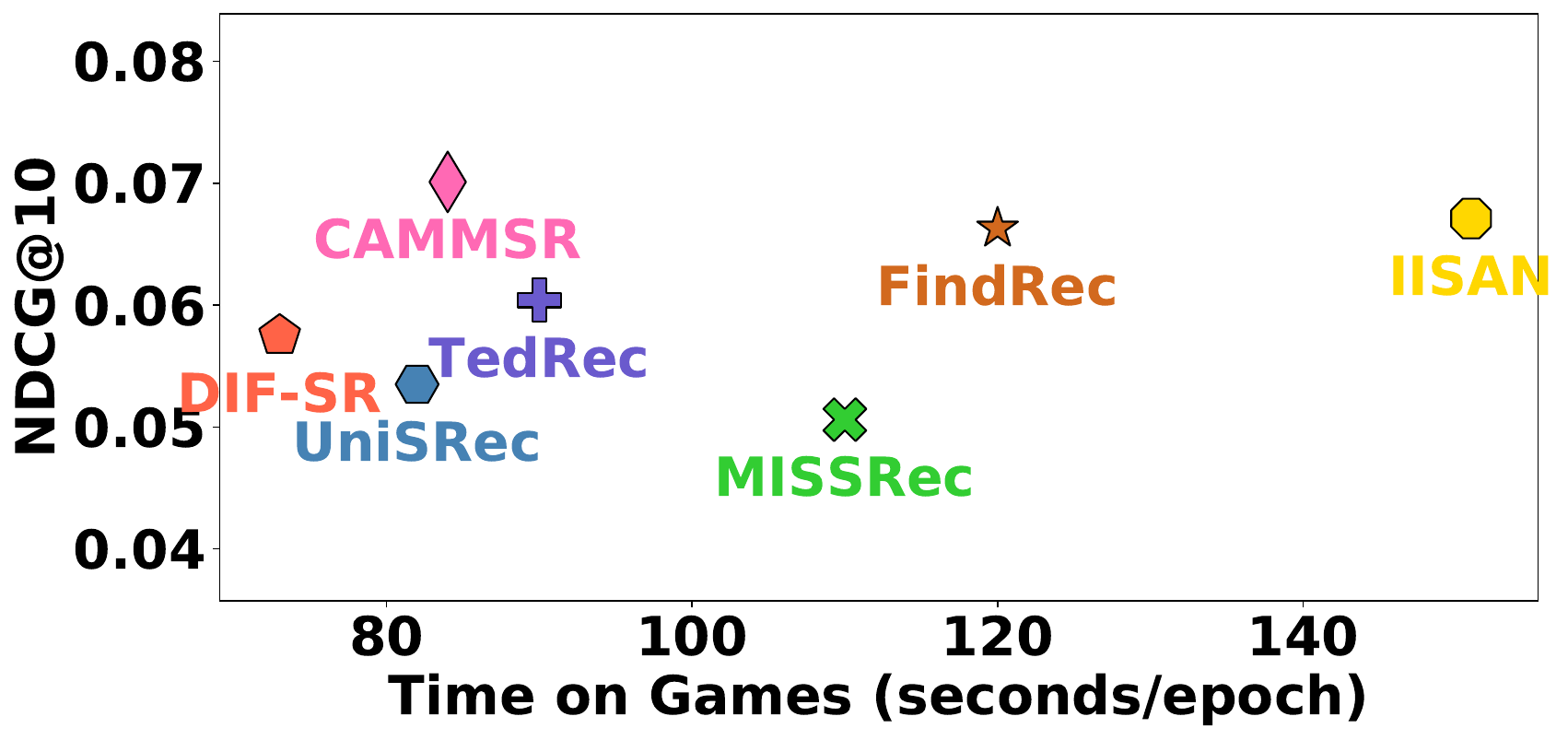}
        }  
    \subfigure[Beauty] {
        \label{fig:c}
        \includegraphics[width=0.8\linewidth]{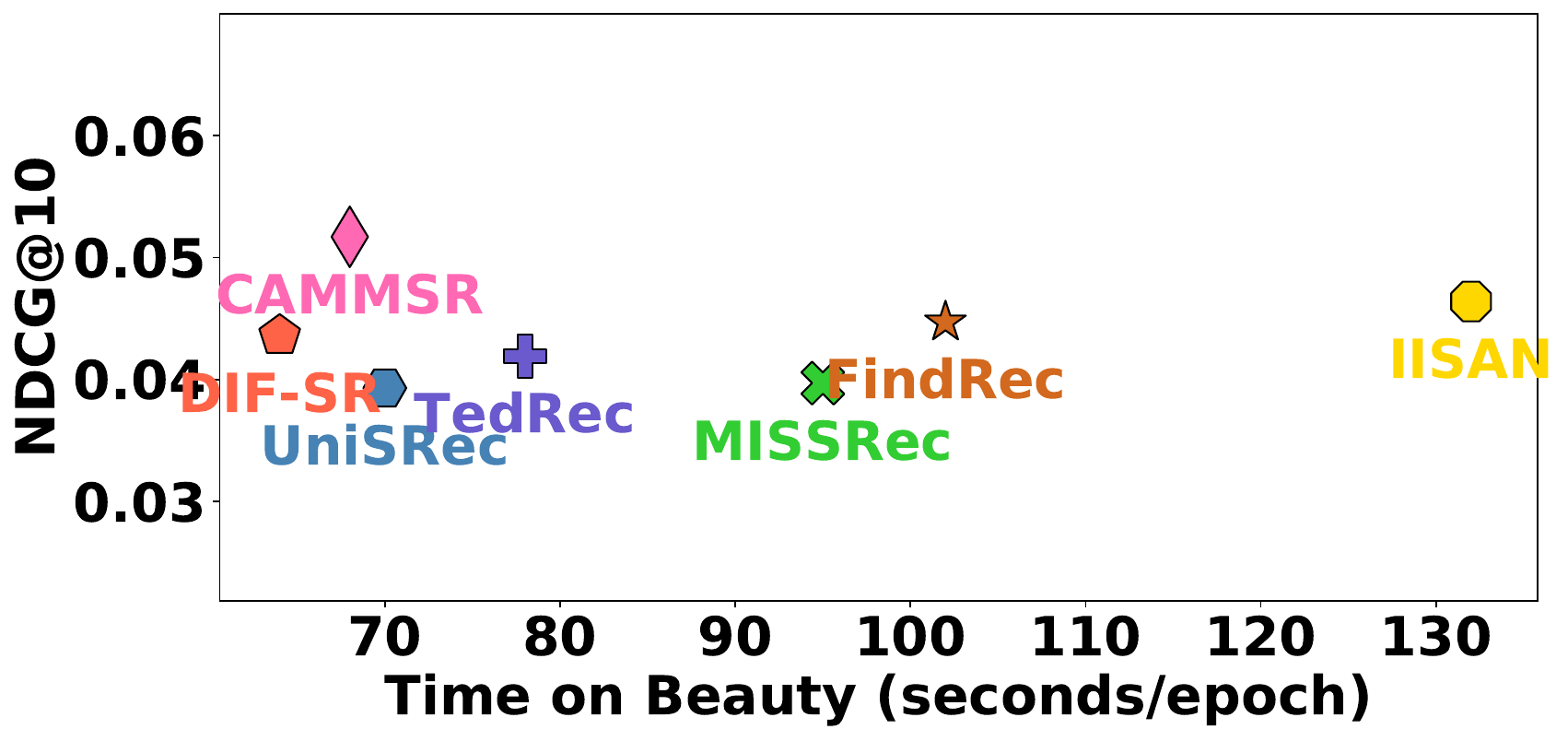}
        }  
    \vskip -0.1in
    \caption{Efficiency study on both Games and Beauty datasets.}   
    \label{fig:efficiency}
    \vskip -0.1in
\end{figure}

\begin{table*}[!ht]
\centering
\vskip -0.1in
\small
 % \vskip -0.05in
% \small
% \tabcolsep=0.1cm
\caption{Efficiency comparison of different baselines across two datasets, including average training time per epoch, number of epochs to converge, total training time, and inference time per epoch (TT/E: Training Time/Epoch, \#E: \#Epoch, TTT: Total Training Time, IT/E: Inference Time/Epoch, N@10: NDCG@10; s: second, m: minute, h: hour).}
% We highlight the optimal model for the TTT metric in \textbf{bold}.
 % \vskip -0.1in
\vskip -0.05in
\label{tab:efficiency}
% \resizebox{1\linewidth}{!}{
\begin{tabular}{ccccccccccc}
\toprule
\multirow{2.5}{*}{\textbf{Model}}&\multicolumn{5}{c}{\textbf{Games}} & \multicolumn{5}{c}{\textbf{Beauty}} \\ \cmidrule{2-6} \cmidrule{7-11} & TT/E$\downarrow$ & {\#E$\downarrow$} & TTT$\downarrow$ & IT/E $\downarrow$ & {N@10$\uparrow$} & TT/E$\downarrow$ & {\#E$\downarrow$} & TTT$\downarrow$ & IT/E $\downarrow$ & {N@10$\uparrow$} \\
\midrule
DIF-SR& 73s& 119& 2h24m47s& 9.2s& 0.0575& 64s& 103& 1h49m52s& 7.9s& 0.0436\\
UniSRec& 82s& 107& 2h26m14s& 11.7s& 0.0535& 70s& 95& 1h50m50s& 9.4s& 0.0393\\
MISSRec& 110s& 89& 2h43m10s& 15.2s& 0.0506& 95s& 74& 1h47m10s& 10.6s& 0.0397\\
IISAN& 151s& 114& 4h46m54s& 23.1s& 0.0671& 132s& 99& 3h37m48s& 19.8s& 0.0464\\ 
TedRec& 90s& 103& 2h34m30s& 11.5s& 0.0604& 78s& 88& 1h54m24s& 10.1s& 0.0419\\
FindRec& 120s& 90& 3h0m0s& 19.8s& 0.0663& 102s& 71& 2h0m42s& 16.0s& 0.0447\\ \midrule
CAMMSR& 84s& 65& \textbf{1h31m0s}& 10.4s& \textbf{0.0701}& 68s& 53& \textbf{1h0m4s}& 8.8s& \textbf{0.0517}\\
\bottomrule
\end{tabular}
\vskip -0.1in
\end{table*}

\subsection{Case Study (RQ6)}
\label{sec:case}
To empirically validate the effectiveness of our proposed category-guided attentive mechanism and its practical implications for passive information management, we conduct a detailed case study using a representative sequence from the Beauty dataset. As shown in Figure~\ref{fig:case study}, we compare the top-1 recommendation generated by CAMMSR against that of IISAN, which achieves the strongest performance among all baseline methods on this dataset. While both models correctly identify the broad product category (CC Cream), CAMMSR uniquely succeeds in recommending the exact ground-truth item that aligns with the user's underlying preference pattern.

This superior performance can be attributed to our category-guided attentive MoE framework, which enables more intelligent passive information management through context-aware modality weighting. Our analysis reveals that the ground-truth item and the second item in the historical sequence both belong to the "Premium Beauty" category. Through the auxiliary category prediction task, CAMMSR learns that for high-value cosmetic products, textual descriptions—containing detailed ingredient lists, usage instructions, and quality assurances—carry greater discriminative power than visual appearances alone. This insight mirrors real-world consumer behavior where purchasers of premium beauty products meticulously examine textual details before making decisions.

Consequently, when processing the target sequence, CAMMSR dynamically amplifies the contribution of textual signals through its attentive gating mechanism, effectively filtering out visually similar but semantically mismatched alternatives. This adaptive modality weighting demonstrates how our framework operates as an advanced passive data management system: it automatically discerns and prioritizes the most relevant informational dimensions based on item category and contextual patterns, thereby reducing users' cognitive burden in navigating complex multimodal content. In contrast, IISAN's static fusion strategy fails to capture these nuanced category-dependent modality preferences, leading to its recommendation of a visually similar but contextually inappropriate alternative.

This case study concretely illustrates how CAMMSR advances multimodal sequential recommendation beyond conventional fusion strategs. By dynamically aligning modality emphasis with category-specific decision patterns, CAMMSR provides a more sophisticated approach to passive information management—automatically surfacing the most relevant content while filtering peripheral signals, ultimately delivering more precise and contextually appropriate recommendations.

\begin{figure}[!h]
    \centering
     \vskip -0.1in
    \subfigure[Hyperparameter $K$] {
        \label{fig:ndcg10_k}
        \includegraphics[width=0.43\linewidth]{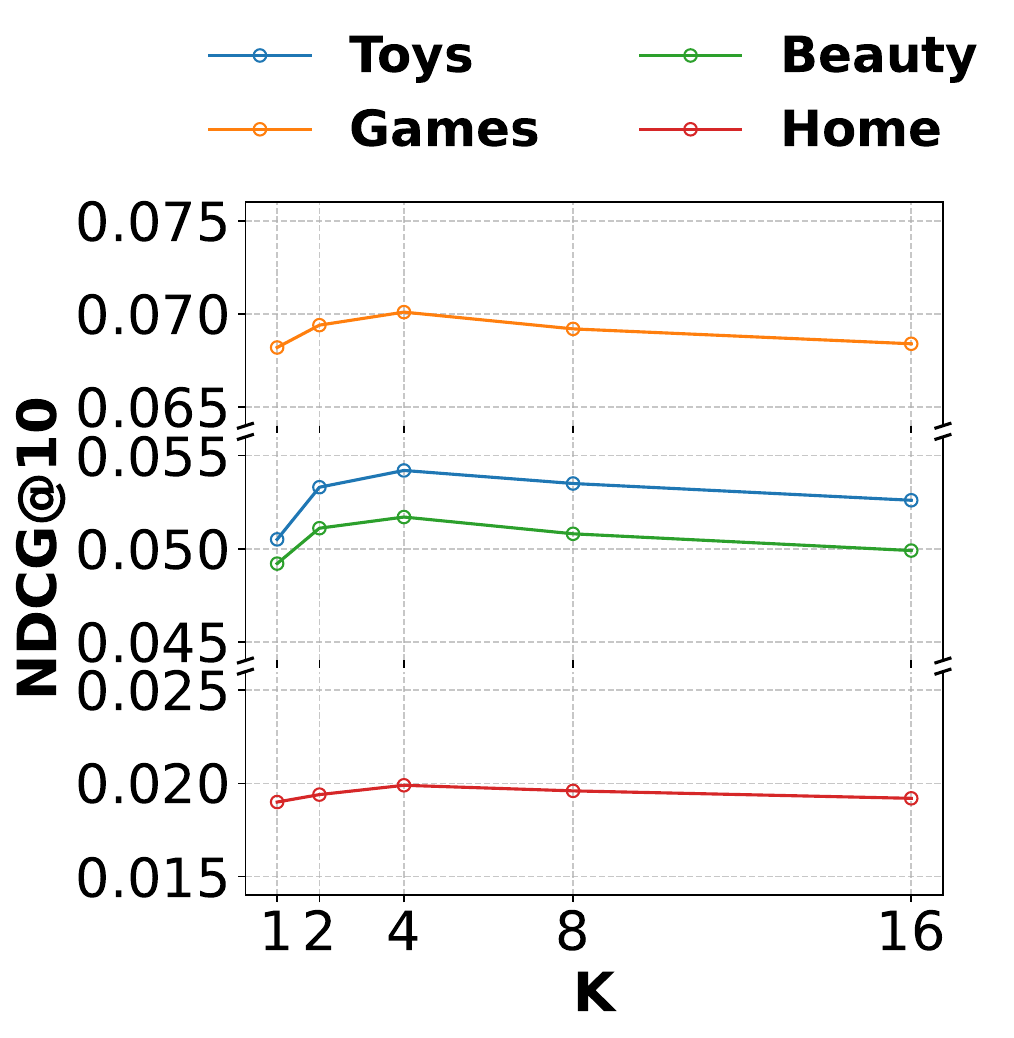}
        }
    \subfigure[Hyperparameter $\lambda$] {
        \label{fig:ndcg10_lambda}
        \includegraphics[width=0.43\linewidth]{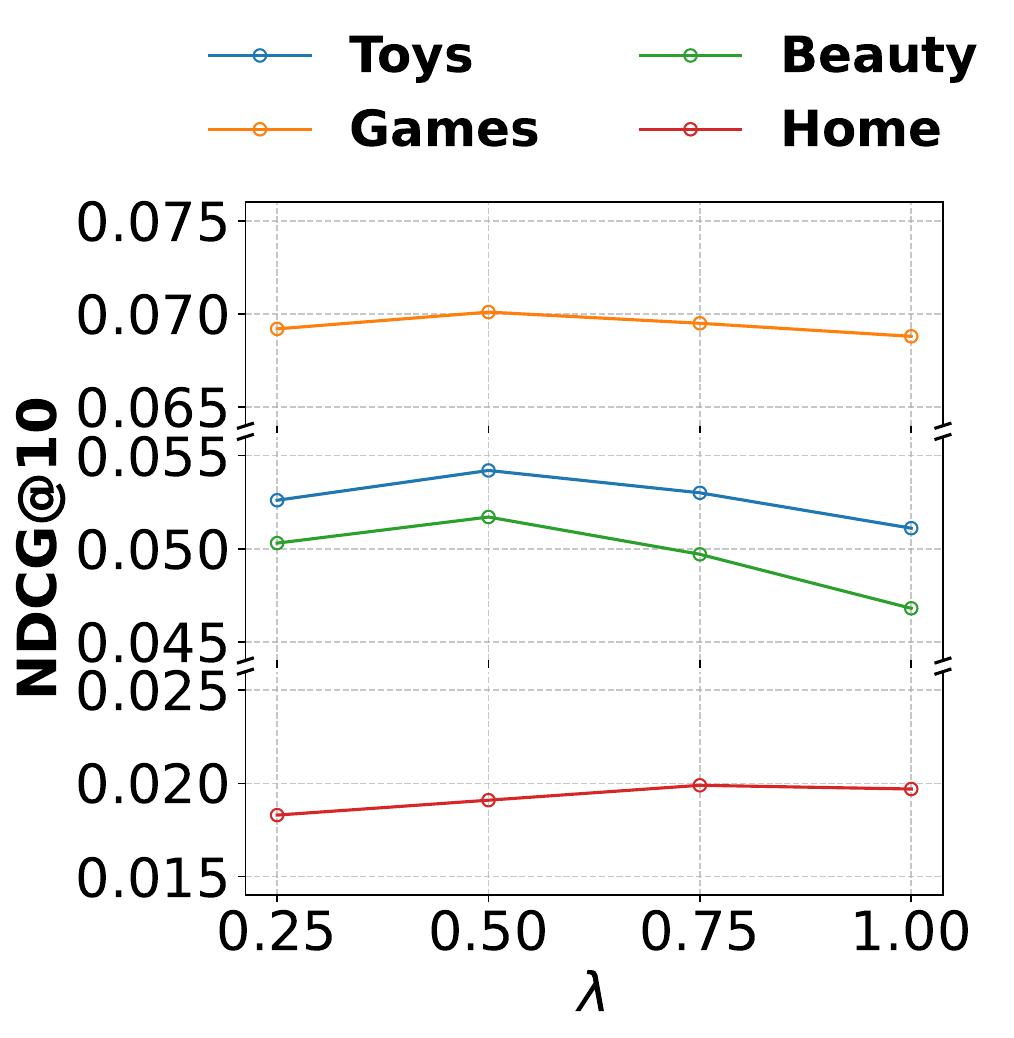}
        }
         \vskip -0.1in
    \caption{Performance comparison $w.r.t.$ key hyper-parameters ($k$ and $\lambda$) in terms of NDCG@10.}
     \vskip -0.1in
    \label{fig:hyper}
\end{figure}

\subsection{Hyper-parameter Analysis (RQ7)}
\label{sec:hyper-parameter}
We further explore the impact of key hyper-parameters on the performance of our CAMMSR. The evaluation results in terms of NDCG@10 are reported in Figure~\ref{fig:hyper}.

\subsection{Performance Comparison $w.r.t.$ $K$:} 
We conduct a detailed study on the influence of the number of experts on model performance, as shown in Figure~\ref{fig:ndcg10_lambda}. The model with 4 experts achieves the best results across all evaluation metrics, outperforming the variant with 1 or 2 experts. This improvement highlights that 1 or 2 experts are insufficient to capture the complex and diverse patterns of cross-modal interactions, whereas 4 experts strike a balance by providing sufficient specialized processing capabilities without imposing significant computational overhead. However, increasing the number of experts to 8 or 16 leads to a decline in performance. This degradation can be attributed to excessive specialization, which introduces redundancy and increases the complexity of the routing mechanism.

\subsection{Performance Comparison $w.r.t.$ $\lambda$:} 
We empirically analyze the effect of the weight hyper-parameter $\lambda$ in our CAMMSR model. As shown in Figure~\ref{fig:ndcg10_k}, $\lambda = 0.25$ is recommended for most datasets, except for the Home dataset, where the optimal setting is $\lambda = 0.5$. We attribute this divergence to the substantially higher sparsity characteristic (99.97\%) of the Home dataset, which amplifies the necessity for enhanced cross-modal alignment through more pronounced contrastive learning signals to bridge modality gaps in data-scarce environments.

This robustness to hyper-parameter selection significantly enhances the practical deployability of our framework in real-world data management scenarios, as it reduces the extensive parameter tuning efforts typically required for multimodal systems while maintaining reliable recommendation quality. The observed stability aligns with the requirements of industrial-scale recommender systems, where operational consistency and minimal maintenance overhead are crucial considerations for long-term deployment sustainability.

\subsection{Category Analysis}
\label{sec:category}
We further explore how category-guided fusion (CG) can be applied in multi-label category, hierarchical category, and no-category scenarios. For multi-label category and hierarchical category scenarios, applying various aggregation strategies (mean/max/min pooling) across multiple differences $d$ proves to be a simple yet effective solution. In none category scenario, utilizing LLMs to construct virtual categories presents a promising solution. Specifically, we employ max pooling for scenarios involving multi-label and hierarchical categories. Additionally, we utilize GPT-4o to generate virtual categories for all items based on textual and visual information. To evaluate this approach, we conduct experiments on the Amazon Baby dataset \cite{zhou2023mmrec,ye2025harnessing}, which includes multi-label categories. For the scenario without existing categories, we remove all original categories and generate virtual categories using the same vocabulary as the original dataset. Moreover, to validate the generalization of CG, we utilize the Microlens dataset \cite{ni2025content} for further evaluation. Specifically, we employ GPT-4o to generate 10 or 20 virtual categories. Table~\ref{tab:category} demonstrates that CG consistently achieves performance improvements across all scenarios. For the Microlens dataset, both 10 and 20 virtual categories deliver consistent and comparable performance improvements compared to the \textbf{w/o CG} variant.

\begin{table}[!t]
 \vskip -0.05in
 % \small
\caption{Performance for CAMMSR and \textbf{w/o CG} variant under multi-label category and none category scenarios.}
\label{tab:category}
 \vskip -0.05in
\centering
\resizebox{\linewidth}{!}{
    \begin{tabular}{ccccc}
     \toprule
         Baby& NDCG@5& NDCG@10& MRR@5& MRR@10\\ \midrule
         Multi-label Category& 64.09& 70.28& 55.29& 61.03\\
         % & \textbf{w/o CG}& & & & \\ \midrule
         None Category& 62.88& 68.05& 53.11& 58.66\\
         \textbf{w/o CG}& 57.13& 62.20& 44.13& 48.70\\ \midrule
         Microlens& NDCG@5& NDCG@10& MRR@5& MRR@10\\ \midrule
         None Category (10)& 61.40& 65.51& 52.59& 56.91\\
         None Category (20)& 61.52& 65.48& 52.45& 57.12\\
         \textbf{w/o CG}& 57.83& 62.21& 46.00& 49.45\\ \bottomrule
    \end{tabular}
    }
    \vskip -0.1in
\end{table}
\section{Conclusion}
In this paper, we presented CAMMSR, a novel Category-guided Attentive Mixture of Experts framework for multimodal sequential recommendation. To address the inherent limitations of static fusion strategies in existing multimodal models, our CAMMSR holistically integrates multimodal content from user behavior sequences and corresponding timestamps, enabling more nuanced and temporally-aware recommendation generation. The core of our framework features a category-guided attentive mixture of experts (CAMoE) module that not only captures specialized item representations from multiple perspectives but also explicitly models the synergistic interactions between different modalities. By dynamically allocating modality weights through an auxiliary category prediction task, our model achieves adaptive fusion of multimodal signals that aligns with both item characteristics and user contextual preferences. Furthermore, the introduced modality swap contrastive learning task effectively strengthens cross-modal alignment and enhances the representation consistency through sequence-level augmentation strategies. Extensive experiments conducted on four real-world datasets demonstrate the effectiveness and efficiency of our proposed CAMMSR. 

Looking forward, CAMMSR establishes a flexible foundation for advancing adaptive multimodal recommendation systems. Its capacity to dynamically integrate multimodal signals while preserving cross-modal consistency supports several meaningful extensions, such as generalizing the category guidance to other contextual dimensions, developing more refined synergy modeling approaches, and adapting the framework to sequential decision tasks outside conventional recommendation settings.

\section{The Use of AI}
We made limited use of large language models (LLMs) for writing assistance only, including grammar correction, style polishing, and table layout/formatting. All proposed changes were manually reviewed and selectively adopted by the authors. All scientific content, ideas, analysis, and conclusions remain entirely our own. The authors take full responsibility for the entire content of this paper, including any errors or inaccuracies that may remain.

\bibliographystyle{IEEEtran}
% \balance
\bibliography{ref}

\end{document}